\documentclass[a4paper,11pt]{article}
\pdfoutput=1
\usepackage{jheppub}
\usepackage{slashed}
\usepackage[T1]{fontenc}
\usepackage{adjustbox}
\usepackage{multirow}
\usepackage{hhline}
\usepackage{tabularx,booktabs,caption}
\usepackage{float}
\usepackage{placeins}
\usepackage{caption}
\usepackage{verbatim}
\usepackage{soul}
\usepackage[utf8]{inputenc}
\usepackage{amsmath,etoolbox}
\AtBeginEnvironment{pmatrix}{\setlength{\arraycolsep}{10pt}}

\usepackage{tikz}
\usepackage{cancel}
\usepackage{graphicx}

\usepackage{xcolor}
\usepackage{hyperref}
\usepackage{dirtytalk}

\newcommand{\beq}{\begin{equation}}
\newcommand{\eeq}{\end{equation}}
\newcommand{\beqn}{\begin{eqnarray}}
\newcommand{\eeqn}{\end{eqnarray}}
\newcommand{\bea}{\begin{eqnarray}}
\newcommand{\eea}{\end{eqnarray}}

\title{\boldmath Extra dimensions with light and heavy neutral leptons: An application to CE$\nu$NS}

\author[1]{Amir N.\ Khan,\note{Corresponding author.}}
\affiliation{ Max-Planck-Institut f\"ur Kernphysik, Postfach
103980, D-69029 Heidelberg, Germany} 

\emailAdd{amir.khan@mpi-hd.mpg.de}

\abstract{We explore the possibility of relating extra dimensions with light and heavy Dirac-type neutral leptons and develop a framework for testing them in various laboratory experiments. The Kaluza-Klein modes in the large extra dimension models of the light neutral leptons could mix with the standard model neutrinos and produce observable effects in the oscillation experiments. We show that the chirality flipping up-scattering processes occurring through either neutrino magnetic dipole moment or the weakly coupled scalar interactions can also produce heavy Kaluza-Klein modes of the corresponding right-handed neutral leptons propagating in one or more extra dimensions. However, to conserve the four-dimensional energy-momentum, their masses must be below the maximum energy of the neutrinos in the initial state. The appreciable size of extra dimensions connected with these heavy neutral leptons can thus affect the cross-sections of these processes. This framework applies to any up-scattering process. Our work here focuses only on its application to the coherent elastic neutrino-nucleus scattering process. We derive constraints on the size of extra dimensions using the COHERENT data in oscillation and up-scattering processes. For model with one large extra dimension for the light neutral leptons, we obtain the limits, $R \sim 3 \ \mu$m (NH) and $R \sim 2.5 \ \mu$m (IH), on the size of extra dimension corresponding to the absolute mass limit, $m_{0} \leq 3 \times 10^{-3}$ eV at 90$\%$ C.L. from the short-baseline oscillations. Using the up-scattering process for heavy neutral leptons, we obtain new parameter spaces between the size of extra dimensions and parameters of the dipole or scalar interactions.}
\begin{document} 
\captionsetup[figure]{labelfont={bf},labelformat={default},labelsep=period,name={FIG.}}
\maketitle

\section{Introduction}\label{sec:intro}
Kaluza and Klein introduced extra dimensions in 1920 to unify gravity with electromagnetism by extending Einstein’s general theory of relativity to five-dimensional space \cite{kaluza1921unitatsproblem, klein1926quantentheorie}. The advent of string theory revived the idea of extra dimensions after a long time. Nowadays, all versions of string theory are consistently formulated in ten space-time dimensions, or eleven in the case of M-theory \citep{Cremmer:1978km, Ginsparg:1987ee, Witten:1995ex, Witten:1996mz, Lykken:1996fj}. It was assumed before that the extra dimensions are compactified over manifolds with radii of the order of Planck length,  $l_p\sim 10^{-33}$ cm, and the corresponding energy scale of the Planck mass, $ M_p \sim 2.4\times 10^{18}$ GeV, where the quantum gravity effects become strong. There was no hope of testing such small $(l_{p})$ and big $(M_{p})$ scales in any experiment. In this paradigm, the Higgs mass requires a large fine-tuning for the vast gap between the electroweak scale $(\sim 1$ TeV) and the Planck scale, or the grand unification scale $(\sim 10^{16}$ GeV), referred to as the hierarchy problem.

Later, the authors of ref. \cite{Antoniadis:1990ew, ArkaniHamed:1998rs, Antoniadis:1998ig, ArkaniHamed:1998nn} introduced the idea of large extra dimensions (LED), suggesting that some of these extra dimensions could be much larger than the Planck length. This model predicts that the extra dimensions' sizes are of the order of a millimeter. However, they remain hidden from the experiments because the LED model assumes that all the standard model (SM) particles are confined to the four-dimensional (4D) Minkowski space, and only gravity propagates in the extra dimensions. In this model, the fundamental scale of quantum gravity can be lowered by about fifteen orders of magnitude less than the Planck scale in 4D, which solves the issue of weak gravity and the hierarchy problem. Soon afterwards, other interesting models, such as AdS/CFT correspondence \cite{Maldacena:1997re}, warped extra dimensions \cite{Randall:1999ee,Randall:1999vf} and universal extra dimensions \cite{Appelquist:2000nn} were introduced. This development made the idea of extra dimensions much more attractive.

Along with the hierarchy problem and weak gravity explanations, the LED model can also explain the finite yet tiny masses of neutrinos \cite{Dienes:1998vg, ArkaniHamed:1998vp}. Several other recent string theory models with extra dimensions and models of weak gravity conjecture also offer a solution to the smallness of massive neutrinos \cite{Arkani-Hamed:2007ryu, Ibanez:2017kvh, Gonzalo:2021zsp, Harada:2022ijm, Montero:2022prj}. In the conventional seesaw mechanisms, very heavy right-handed neutrino $\footnote{We will interchangeably use the words 'right-handed neutrinos' and 'heavy neutral leptons' throughout the text. We will call them 'light' or 'heavy' bulk neutrinos when they are assumed to propagate in extra dimensions.}$ partners are necessary for generating the small Majorana-type neutrino masses at a much higher energy scale at the expense of a large gap between the electroweak scale and the higher energy scale. Alternatively, the LED model assumes that gravity and the SM singlet degrees of freedom, in our case the right-handed neutrinos, propagate in bulk. In contrast, the SM particles including the left-handed neutrinos, reside on the 4D brane in the LED model, which is consistent with all the SM predictions up to the TeV scale. The large volume of the extra dimensions suppresses the wave function of the singlet degrees of freedom on the brane, which gives rise to the small Dirac-type neutrino masses in the lowest Kaluza-Klein (kk) modes of higher dimensions. Since there is no constraint on the right-handed neutrino masses, any of them within the mass range from sub-eV to the Planck mass can contribute. The light sterile neutrino can be such a candidate which can be decomposed into a tower of kk excitation that can mix with the active neutrinos and therefore affects the standard neutrino oscillations \cite{Dvali:1999cn,Mohapatra:1999af,Mohapatra:1999zd,Ioannisian:1999cw,Mohapatra:2000wn,Barbieri:2000mg,DeGouvea:2001mz,Davoudiasl:2002fq,Bhattacharyya:2002vf,Machado:2011jt,Machado:2011kt,BastoGonzalez:2012me,Basto-Gonzalez:2012nel,Esmaili:2014esa,Rodejohann:2014eka,DiIura:2014csa,Adamson:2016yvy,Berryman:2016szd,Carena:2017qhd,Becker:2017ssz,Basto-Gonzalez:2021aus}.  

Like the light kk sterile neutrino states (light bulk neutrinos), which mix with the active neutrinos, we assume that the heavy right-handed partners (heavy bulk neutrinos) also propagate in extra dimensions. In laboratory experiments, these can be produced in the chirality flipping scattering processes through either the dipole interactions \cite{ArkaniHamed:1998vp, McLaughlin:1999br,Balantekin:2013sda,Antusch:2017ebe, Magill:2018jla,Shoemaker:2018vii,Bolton:2021pey,Miranda:2019wdy,Brdar:2020quo,Kim:2021lun,Drewes:2022akb,Giffin:2022rei,Calderon:2022alb,Abdullahi:2022jlv,Hernandez:2022ivz} or scalar interactions \citep{Farzan:2018gtr, Brdar:2018qqj}, often called the up-scattering processes. The heavy bulk neutrinos produced in the final state can be related to the extra dimensions at all energy scales. The only constraint that restricts the size of their masses comes from the kinematics of the process. Their masses must be less than the energy of the incoming neutrinos so that the 4D energy-momentum is conserved. Here, we will focus only on the coherent neutrino-nucleus elastic scattering (CE$\nu$NS) process, though the framework developed here can be implemented for the high energy collider experiments which are sensitive to larger mass ranges of the heavy bulk neutrinos such as LHC \cite{Alimena:2019zri}, FASER \cite{FASER:2018bac}, MATSULA \cite{Curtin:2018mvb}, ShiP  \cite{SHiP:2021nfo, Alekhin:2015byh}, Bell II \cite{Belle-II:2018jsg}, FCC \cite{FCC:2018byv}. Some earlier works have studied extra dimensions in high energy colliders \citep{Giudice:1998ck,Han:1998sg,Rizzo:1998fm,deGiorgi:2021xvm}.

Most of the neutrino oscillation experimental results agree well with the three massive neutrinos mixing model, however, there are some experimental anomalies as observed by Gallium experiments \cite{Hampel1998FinalRO,SAGE:2009eeu,Barinov:2022wfh}, LSND \cite{LSND:1995lje, LSND:2001aii}, MiniBoone \cite{MiniBooNE:2007uho,MiniBooNE:2012maf,mini:prl} and the reactor anomalies \cite{Dentler:2017tkw, Schoppmann:2021ywi}. The results from these experiments call for the existence of light sterile neutrinos \cite{Abazajian:2012ys, Dasgupta:2021ies, Acero:2022wqg}. The phenomenology of such light sterile neutrinos is often done in four space-time dimensions (4D). However, this picture changes when the sterile neutrinos are considered to be propagating in the higher dimensions \cite{Barbieri:2000mg,Ioannisian:1999cw,Davoudiasl:2002fq, BastoGonzalez:2012me, DiIura:2014csa,Bhattacharyya:2002vf,Becker:2017ssz,Adamson:2016yvy,Mohapatra:1999zd,Carena:2017qhd,Machado:2011jt,Machado:2011kt,DeGouvea:2001mz,Esmaili:2014esa,Rodejohann:2014eka,Berryman:2016szd,Basto-Gonzalez:2021aus}. It becomes possible to constrain the size and the absolute mass of the lightest neutrino for the normal and inverted mass hierarchies. All these anomalies are related to neutrino production through the charged-current processes. The recently observed neutral CE$\nu$NS process by the COHERENT experiment \cite{Akimov:2021dab, Akimov:2017ade} and by reactor experiments \cite{Aguilar-Arevalo:2019jlr, Colaresi:2022obx} also provide an alternative way of the detection of light sterile neutrinos \cite{Drukier:1984vhf,Anderson:2012pn,Dutta:2015nlo,Blanco:2019vyp,Miranda:2020syh,Berryman:2019nvr,Li:2020lba}. We will investigate this possibility of testing the light sterile LED model with the CE$\nu$NS process. The CE$\nu$NS occurs when the momentum transfer from the incoming neutrino to the target nuclei is small enough that the condition $q^{2}r^2<1$, where $r $ is the nuclear radius, and $q$ is four-momentum transfer, is satisfied.
In the laboratory experiments, for average heavy nuclei, this can occur for the
neutrinos of energy of about 50 MeV and below \citep{Aguilar-Arevalo:2019jlr, Drukier:1983gj, Scholberg:2005qs, Wong:2008vk}. The sensitivity of this process to a variety of new physics has been widely studied before \cite{ deNiverville:2015mwa, Lindner:2016wff,Bauer:2018onh,Heeck:2018nzc,Billard:2018jnl,Denton:2018xmq, Aguilar-Arevalo:2019zme,Bischer:2019ttk,Dutta:2019eml,Khan:2019cvi,Tomalak:2020zfh, Suliga:2020jfa,Khan:2021wzy, Coloma:2022avw, Corona:2022wlb, delaVega:2021wpx, Cerdeno:2021cdz, Alikhanov:2021dhb,CONUS:2022qbb, Khan:2022not, Khan:2022jnd, Abdullah:2022zue}.

In this work, we consider three cases for the bulk neutrinos in connection to the CE$\nu$NS process. In the $first$ case, when the light bulk neutrinos mix with the active neutrinos, we will look for the deficit of the given SM flavor in the detector after short-baseline oscillations. In the $second$ case, the heavy right-handed neutrinos are produced in the CE$\nu$NS process through the dipole interactions due to the neutrino magnetic moment. Finally, in the $third$ case, we will consider that the scalar interactions produce the heavy bulk neutrinos by flipping the chirality of the SM neutrino in the initial state. In the first case, the mass of the bulk neutrino is determined by the flavor oscillations. In contrast, in the second and third cases, the mass of bulk neutrinos depends on the energy of incoming neutrinos and the kinematics of the CE$\nu$NS process. Using the COHERENT data, we will derive limits on the size of extra dimensions for the three cases in different scenarios having different numbers of extra dimensions and kk-modes that are kinetically accessible.

The rest of the paper is organized as follows. First, we review the general formalism, calculate short-baseline oscillation probabilities and cross-sections for magnetic dipole and scalar interactions and then introduce the frameworks for extra dimensions in Sec. \ref{sec:form}. Next, in Sec. \ref{sec:analysis}, we discuss the event rate calculation in the three frameworks for the CE$\nu$NS. Next, we discuss the phenomenology and results in Sec. \ref{sec::Phenomenology}. Finally, we discuss the conclusion and future outlook in Sec. \ref{sec:concl}.

\section{\label{sec:form} Formalism}

To see how the large extra dimensions model can explain the smallness of the neutrinos masses, we consider a volume configuration in ‘$d$’ large extra dimensions with $R_i \ (i = 1, 2, 3....d)$ as the size of $ith$ large dimension. For simplicity we assume a symmetric size ($ R_{1}=R_{2}=R_{3}....= R_{d}=R$) of the $d$-large extra dimensions. The LED model suggests that the 4D Planck mass $(M_p)$ is suppressed by the $d$-dimensional volume factor, $(2\pi R)^d$, according to the following relation \cite{Antoniadis:1990ew,ArkaniHamed:1998rs,Antoniadis:1998ig, ArkaniHamed:1998nn},
\newline
\begin{equation}
M_{P}^{2}=M_{G}^{2+d}(2\pi R)^{d}, 
\label{eq:plkscale}
\end{equation}%
where ‘$M_{G}$’ is the (lowered) true gravitational scale.
The true scale resides in between the electroweak scale and the
$M_{P}$ in $(4D+d)$ dimensional space. For $M_{G} \sim 1$ TeV, the $d = 1$ case is
phenomenologically excluded since the size of the extra dimension is astronomically large, while only $d\geqslant 2$ with a maximum LED size of $\sim 1$ mm are acceptable \cite{ParticleDataGroup:2020ssz}.

The smallness of the active neutrinos can be explained in a model-independent way using the same volume suppression factor of eq. (\ref{eq:plkscale}). The Higgs-neutrino interaction can, therefore, reads as \cite{ArkaniHamed:1998vp}
\newline
\begin{equation}
\mathcal{L} \supset h(M_{G}/M_{P})H%
\overline{\nu }_{L}\nu _{R}, 
\label{eq:higgs}
\end{equation}%
where ‘$H$’ is the Higgs doublet, ‘$h$’ is the Yukawa coupling, ‘$\nu _{L}$’ is brane neutrino, ‘$\nu _{R}$’ is the bulk neutrinos propagating in the $d$-LED. After the electroweak symmetry breaking, the active neutrino mass is 
\newline
\begin{equation}
m_{D}=hvM_{G}/M_{p}, 
\label{eq:mass}
\end{equation}%
where $v=246$ GeV is the vacuum expectation value of $H$. For $h=1$ and $M_{p} \sim 10$ TeV, the brane neutrino mass is $m_{D} \sim 10^{-4}$ eV. 

The bulk neutrinos could be light for one large extra dimension and the $0th$ kk-mode. However, they can be heavy in higher dimensions and with higher kk-modes. The heavier kk-modes are the Dirac partners of the corresponding left-handed heavier kk-modes in one or more extra dimensions. The first case is relevant for the neutrino oscillation, where the SM neutrinos oscillate into the lightest kk-modes, and their effects can be observed in oscillation experiments. In contrast, the heavier bulk neutrinos can be tested in the scattering processes when the SM neutrinos change their chirality and produce massive right-handed neutrinos in the final state. Next, we will discuss how to relate the light and heavy bulk neutrinos with the extra dimensions in the three cases.
\subsection{Light Dirac bulk neutrinos}
 The most straightforward and phenomenologically attractive model relevant for the neutrino oscillations and the absolute mass experiments is the (4D+1)-dimensional LED model. In this model, only one extra dimension is assumed to be much larger than the others, affecting the neutrino oscillation pattern. At the same time, all the others are assumed to be too small to be accessed by these experiments \cite{Dvali:1999cn, Mohapatra:1999af,Mohapatra:1999zd, Ioannisian:1999cw,Mohapatra:2000wn, Barbieri:2000mg, DeGouvea:2001mz,Davoudiasl:2002fq, Bhattacharyya:2002vf}. Presumably, some fraction of the active neutrinos can efficiently oscillate to the light bulk neutrinos for a small mass-squared difference. The standard oscillations pattern and the effects should appear in the resulting energy spectrum. The phenomenology of this model has consistently been studied. Constraints on the LED size were derived for several short-, long-baseline, and neutrino absolute mass experiments \cite{Machado:2011jt, Machado:2011kt, BastoGonzalez:2012me, Basto-Gonzalez:2012nel, Esmaili:2014esa, Rodejohann:2014eka,DiIura:2014csa, Adamson:2016yvy,Berryman:2016szd, Carena:2017qhd,Becker:2017ssz, Basto-Gonzalez:2021aus}. We will examine this possibility here with the short-baseline CE$\nu$NS experiment, derive the constraints, and compare them with the existing ones.

In this model, the light bulk neutrinos couple to the left-handed 4D brane neutrinos \footnote{We will interchangeably use the words ‘SM neutrinos’ and ‘brane neutrinos’ throughout the text.} through the SM Higgs field, leading to the Yukawa couplings according to eq. (\ref{eq:higgs}). This coupling generates mixings between the brane neutrinos and the kk-modes of the light bulk neutrinos in extra dimensions. After the electroweak symmetry breaking, one can write the resulting mass terms and the SM charged-current interaction terms in the 4D space-time as \cite{Davoudiasl:2002fq}
\begin{align}
\hskip -0.5cm
\mathcal{L} &\supset
\sum_{\alpha,\beta}m_{\alpha\beta}^{D}\left[\overline{\nu}_{\alpha
    L}^{\left(0\right)}\,\nu_{\beta R}^{\left(0\right)}+\sqrt{2}\,
  \sum_{n=1}^{\infty}\overline{\nu}_{\alpha
    L}^{\left(0\right)}\,\nu_{\beta
    R}^{\left(n\right)}\right]  
 +  \sum_{\alpha}\sum_{n=1}^{\infty}\displaystyle
\frac{n}{R}\, \overline{\nu}_{\alpha L}^{\left(n\right)} \,
\nu_{\alpha R}^{\left(n\right)} \notag  \\  & \quad 
+\frac{g}{\sqrt{2}}\,\sum_{\alpha} \,
\overline{l_\alpha}\gamma^{\mu}\left(1-\gamma_{5}\right)\nu_{\alpha}^{\left(0\right)}\,
W_{\mu}+\mbox{h.c.},
\label{lagran}
\end{align}
where $\alpha,\beta = e,\mu,\tau$ are the flavor indices, $n=0, 1,2,3,...,\infty$ are the kk-modes, $m_{\alpha \beta}^{D}$ is the Dirac mass
matrix, $\nu^{(0)}_{\alpha R}$, $\nu^{(n)}_{\alpha R}$ and
$\nu^{(n)}_{\alpha L}$ are the linear combinations of the bulk fermion
fields that couple to the brane neutrinos, $\nu^{(0)}_{\alpha L}$. 

To diagonalize the mass term, we change to the mass basis using the following redefinitions of the fields, 
\begin{eqnarray}
\nu_{\alpha L}^{(n)} &=& \sum_{i=1}^3 U'_{\alpha i } \nu_{L}^{' i(n)},\  \quad n=0,1,2,\ldots, \infty\,
\label{lhnus}
\end{eqnarray}
\begin{eqnarray}
\nu_{\alpha R}^{(n)} &=& \sum_{i=1}^3 U'_{\alpha i } \nu_{R}^{' i(n)},\ \quad n=0,1,2,\ldots, \infty\,, 
\label{rhnus}
\end{eqnarray}
where ‘$U'_{\alpha i }$’ is the 3 $\times$ 3 matrix which diagonalizes the Dirac mass matrix and for $n = 0$, the $0th$ kk-mode, $U'_{\alpha i } \equiv U_{\alpha i}$ in eq. (\ref{lhnus}) such that diag$(m_1,m_2,m_3) = U'{}^\dagger m^D U$, where $U$ is the standard leptonic mixing matrix. We can write the mass term in a more explicit form as, 
\begin{equation}
\mathcal{L}=\sum_{i=1}^3 \overline{\nu}_L^{' i} M^i\nu_R^{' i} + \mbox{h.c.} \,,
\end{equation}
where ${\nu_{L}^i}^T \equiv(\nu_{L\\}^{i (0)}, \nu_{R}^{i (1)}, \nu_{R}^{i (2)} \ldots)$ 
and 
$\nu_{R}^i=(\nu_{R}^{i (0)}, \nu_{R}^{i (1)}, \nu_{R}^{i (2)}, \ldots)$ are infinite-dimensional vectors, and the mass matrix, $M_{i}$, for $(n+1)$ kk states looks as follows,
\begin{equation}
M_i = \frac{1}{R}
\begin{pmatrix}
\xi_{i}/\sqrt{2} & \xi_{i}   & \xi_{i}  & \ldots\\
0                & 1         & 0        & \ldots\\
0                & 0         & 2        & \ldots\\
\vdots           & \vdots    & \vdots   & \ddots,
\end{pmatrix}
\end{equation}
where $\xi_{i} \equiv \sqrt{2}m_{i} R$. Finally, we require two matrices $L$ and $R$ of infinite dimensions to diagonalize $L^{i \dagger} M^i R^i$ while the corresponding mass eigenstates are given by $\nu_L^i = L_i^\dagger \nu_L^i{}'$ and $\nu_R^i = R_i^\dagger \nu_R^i{}'$. The standard left-handed massive neutrinos flavor eigenstates, $\nu_L^\alpha$, in terms of the bulk neutrinos mass eigenstates with `n' kk-modes, $\nu_L^{i (n)}$, are therefore given by,
\begin{equation}
    \nu_L^{\alpha (0)} = \sum_{i=1}^3 U_{\alpha i} \nu_L^{' i(n)}{}\, = \sum_{i=1}^3 \sum_{n=0}^\infty U_{\alpha i} L_i^{0 n}\nu_L^{i(n)}{}\,,
    \label{eq:L0n}
\end{equation}

By diagonalizing the quantity $M^{\dagger}M$, one can easily obtain the corresponding eigenvectors \cite{Dvali:1999cn,Davoudiasl:2002fq,Machado:2011jt}. The $L_{i}^{0 n}$ can therefore be obtained as \cite{Davoudiasl:2002fq},
\begin{align}
\hskip -0.5cm
\left(L_{i}^{0n} \right)^{2} &= \frac{2}{1+\pi^{2}\xi_{i}^{2}/2+2\lambda_{i}^{(n)2}/\xi_{i}^{2}},
\end{align} 
where ‘$\lambda_{i}^{(n)}$’ are the eigenvalues of the corresponding Hermitian matrix $R^2 M^{\dagger}M$ which can be found from the eigenvalue equation, $\lambda_j^{(n)}= m_{i}^{(n)}R$, by solving its characteristic equation det$(R^2 M^{\dagger}M-\lambda_j^{(n)} I)=0$. It leads to the following transcendental equation \cite{Davoudiasl:2002fq},

\begin{align}
\hskip -0.5cm
\lambda^{(n)}_{i}-\frac{\pi}{2} \xi_{i}^2 \cot(\pi \lambda^{(n)}_{i}) &= 0.
\end{align}
For small $\xi_{i} (m_{i}^D R<<1)$, the eigen-values, up to the leading terms, are

\begin{eqnarray}
\lambda_{i}^{(n)} = \left \{ \begin{array}{cc}
 \frac{1}{\sqrt{2}} \xi_{i} - \frac{1}{12 \sqrt{2}}\pi^{2} \xi_{i}^{3} +\mathcal{O}(\xi_{i}^5), & n=0 \\
n + \frac{1}{2n} \xi_{i}^{2}- \frac{1}{4 n^{3}}\xi_{i}^{4} +\mathcal{O}(\xi_{i}^6),  & \ \ n=1,2,3 ...
\end{array}
\right.
\label{eq:eigenvalue}
\end{eqnarray}
Using eq. (\ref{eq:eigenvalue}), one easily derive the approximate mixing matrix elements of eq. (\ref{eq:L0n}) as in the following

\begin{eqnarray}
L_i^{0n} = \left \{ \begin{array}{cc}
 1-\frac{\pi^{2}\xi_{i}^{2}}{12}+\frac{7\pi^{4}\xi_{i}^{4}}{1440}+\mathcal{O}(\xi_{i}^6), & n=0 \\
\frac{\xi_{i}}{n}-\frac{3\xi_{i}^3}{4 n^3}-\frac{9\xi_{i}^5}{32 n^5}+\mathcal{O}(\xi_{i}^7),  & \ \ n=1,2,3 ... 
\end{array}
\right.
\label{eq:impli}
\end{eqnarray}
Next, we derive the oscillation probabilities for the CE$\nu$NS using the above setup.
\subsubsection{Light bulk neutrinos via oscillations in CE$\nu$NS}
In terms of the eigenvalues of eq. (\ref{eq:eigenvalue}) and the mixing matrix elements of eq. (\ref{eq:impli}), the amplitude ‘${\cal{A}}$’ of neutrino flavor transition at a distance ‘$L$’ from the neutrino production can be written as,
\begin{eqnarray}
{\cal{A}}_{\nu_\alpha^{(0)} \nu_\beta^{(0)}} (L) & = & \displaystyle
\sum_{i=1}^{3}\sum_{n=0}^{\infty} U_{\alpha i} U_{\beta i}^{*}
\left( L_i^{0n}\right)^2 \times \exp \left(i\frac{\lambda_i^{(n)2}L}{2E_{\nu}R^2} \right)\, ,
\label{eq:amplitude}
\end{eqnarray}
where ‘$U_{\alpha i}$’ are the elements of standard $3\times 3$ leptonic mixing matrix, ‘$E_{\nu}$’ is the neutrino energy, ‘$\lambda_i^{(n)}/R$’ are the masses of the corresponding neutrino states, ‘$\nu_L^i$’. The transition amplitude of eq. (\ref{eq:amplitude}) simplifies for the case of survival probability by splitting the terms $n = 0$ terms and those $n > 0$ modes and then using the approximation, $m_{i}^D R<<1$, and the eq. (\ref{eq:eigenvalue}). Given the general precision level of the current neutrino experiments and the approximation, $\xi_i<<1$, it is sufficient to consider the second-order terms in ‘$\xi_i$’ and first-order terms in the phases both for the $n=0$ and $n>0$ kk-modes. It leads to the following simple form of the oscillation amplitude for the first six kk-modes, 

\begin{align}
{\cal{A}}_{\nu_\alpha^{(0)} \nu_\alpha^{(0)}} (L) =
\sum_{i=1}^{3} |U_{\alpha i}|^{2}
\left[\left(1-\frac{\pi^{2}\xi_{i}^{2}}{12}\right)^2\exp \left(i\frac{\xi_i^{2}L}{4E_{\nu}R^2} \right)+
\sum_{n=1}^{5}
\left(\frac{\xi_{i}^2}{n^2}\right)\exp \left(i\frac{(\xi_{i}^{2}+n^2)L}{2E_{\nu}R^2} \right)\right],\
\label{eq:amplitudesur1}
\end{align}
One can write this in a more explicit form as,
\begin{align}
{\cal{A}}_{\nu_\alpha^{(0)} \nu_\alpha^{(0)}} (L) =
\sum_{i=1}^{3} |U_{\alpha i}|^{2}
\left[\left(1-\frac{\pi^{2}m_{i}^2 R^{2}}{6}\right)^2\exp \left(i\frac{m_{i}^2 L}{2E_{\nu}} \right)+
\sum_{n=1}^{5}
\left(\frac{2m_{i}^2 R^{2}}{n^2}\right)\exp \left(i\frac{(2m_{i}^2 R^{2}+n^2)L}{2E_{\nu}R^2} \right)\right]\, ,
\label{eq:amplitudesur2}
\end{align}
and the oscillation probability is given by
\begin{align}
P_{\nu_\alpha^{(0)}\nu_\alpha^{(0)}} = &  \displaystyle
\left|{\cal{A}}_{\nu_\alpha^{(0)} \nu_\alpha^{(0)}} (L) \right|^{2}.
\label{eq:prob}
\end{align}

\begin{figure} \centering
\includegraphics[width=1.05\textwidth,height=0.56\textwidth]{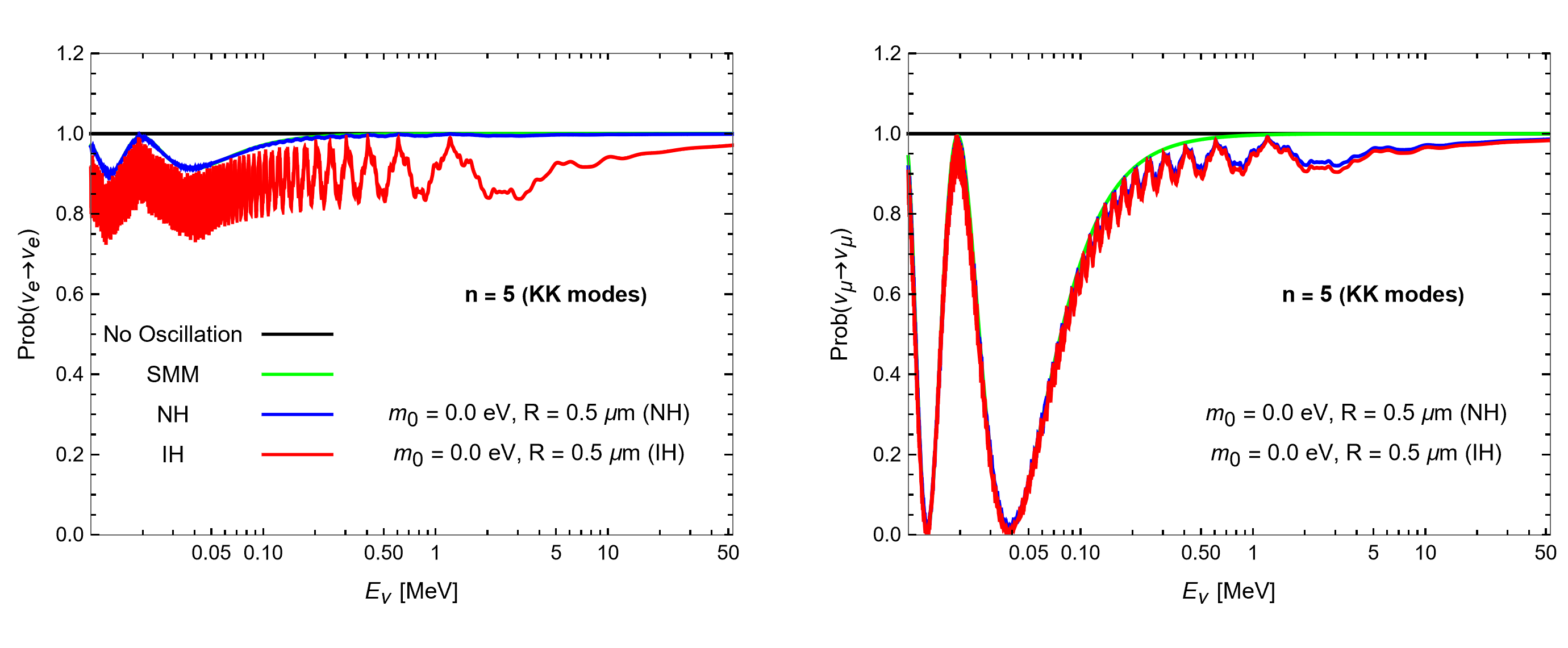}
\caption{Oscillation probabilities relevant for the COHERENT set up with normal and inverted hierarchies. We have used the oscillation parameter values, $\Delta m^2_{ij} \equiv 7.6 \times 10^{-5}$ eV$^{2}$, $\sin^2{2\theta_{12}} =0.31, |\Delta m_{31}^2|=2.4\times 10^{-3} $eV$^2$ for this plot.}
\label{probb}
\end{figure}

Using eq. (\ref{eq:prob}), we calculate the neutrino survival probabilities of ‘$\nu_{e}$’ and ‘$\nu_{\mu}$’ flavors relevant for the COHERENT setup with a baseline of 19.3 meters. For simplicity, we assume CPT invariance and treat ‘$\nu_{\mu}$’ and ‘$\overline{\nu}_{\mu}$’ the same while we further assume that ‘$\nu_{\mu}$’ and ‘$\nu_{e}$’ flavor oscillate to the same bulk neutrino flavors at the detection. The resultant probabilities as a function of the neutrino energy are shown in Fig. \ref{probb}, for the case of normal hierarchy (NH) with $m_3 > m_2 > m_1 \equiv m_0$ and inverted hierarchy (IH) with $m_2> m_1 > m_3 \equiv m_0$ where the SM neutrino mass squared differences are defined as $\Delta m^2_{ij} \equiv m^2_i - m^2_j$ ($i,j = 1,2,3$). We use the following values of the oscillation parameters for the analysis, $\Delta m^2_{ij} \equiv 7.6 \times 10^{-5}$ eV$^{2}$, $\sin^2{2\theta_{12}} =0.31, |\Delta m_{31}^2|=2.4\times 10^{-3} $eV$^2$ \cite{ParticleDataGroup:2020ssz}. As shown in the figure, we consider the bulk neutrinos up to 5 kk-modes for Fig. \ref{probb} and the rest of the analysis. For these plots, we assume $m_0=0.0$ eV and $R= 0.5\mu m$ for all the cases.
In these figures, we have shown the no oscillations (black curves) and standard mixing model (SMM) oscillations (green curves) case for comparison, where we also take $m_{0}=0$ and $R=0$.

In fig. \ref{probb}, for both survival probabilities, $P(\nu_e\rightarrow \nu_e)$ and $P(\nu_\mu\rightarrow \nu_\mu)$, the fast oscillations occur below 1 MeV due to the ‘$E^{-1}_\nu$’ dependence of the oscillation phases, both for the standard phase term and for the bulk mode terms. However, as one can see, these oscillations are stronger for the IH case. Also, the green curves correspond to the standard oscillation case, with $m_{0} = 0$ and $R = 0$, the fast oscillations do not occur. They start appearing due to the finite size of the extra dimension. Note that the different oscillation patterns and mismatches in the minima of NH and IH occur for the same values of ‘$m_0$’ and ‘$R$’. While the fast oscillations region below 1 MeV is not significant for the COHERENT experiment, it could be important for future CE$\nu$NS experiments with higher sensitivity to the lower recoils or reactor neutrinos \cite{CONUS:2022qbb}.
\subsection{Heavy Dirac bulk neutrinos}
Suppose that any chirality changing process produces the Dirac-type heavy neutral leptons, which propagate in the `$d$' extra dimensions with `$n$' kk-modes and `$p_{i}$' momenta of the kk-modes in the higher dimensions. In this case, it is straightforward to use the energy-momentum conservation in $4D+d$-dimensions and derive a general relation between the mass of the bulk neutrinos, size, and number of large extra dimensions as given in the following,
\begin{equation}
M_{d}^{2}= m_{\nu _{R}}^{2} + \sum_{i=1...d}{\frac{n_{i}^{2}}{R^{2}}}, 
\label{massD}
\end{equation}%

\noindent where ‘$m_{\nu_{R}}$’ is the mass of the right-handed neutrinos in the 4D space-time dimensions, $n_{i} = 0,1,2...\infty$ is number of 
kk-modes for any $ith$ large dimension. Eq. (\ref{massD}) represents the momenta of kk-modes in higher dimensions, but from a 4D point of view, they contribute to the mass of the right-handed neutrinos. As mentioned, the chirality flipping processes can produce these heavy right-handed neutrinos in the final state of the process. The strength of kk-mode or the size of the extra dimensions strongly depends on the kinematical limit of an experiment. From the 4D point of view, the total mass of the outgoing neutrinos must be less than the maximum neutrino energies. The exact form of the mixings of the right-handed bulk neutrinos is given in Eqs. (\ref{lagran}) and (\ref{rhnus}). The neutrino flavor and mass mixings are irrelevant in scattering processes. Therefore, the SM neutrinos of a specific flavor can be related to all kk-modes of heavy right-handed bulk neutrino through the neutrino magnetic moment interactions induced at the quantum loop level or through the weakly coupled scalar interactions.

Interestingly, this idea can be applied to both low energy scattering experiments, in our case to CE$\nu$NS, and to high energy scattering experiments such as LHC \cite{Alimena:2019zri}, FASER \cite{FASER:2018bac}, MATSULA \cite{Curtin:2018mvb}, ShiP  \cite{SHiP:2021nfo,Alekhin:2015byh}, Bell II \cite{Belle-II:2018jsg}, FCC \cite{FCC:2018byv}. The constraints derived from the heavy right-handed neutrino masses produced through the dipole interactions in refs. \cite{Magill:2018jla,Shoemaker:2018vii,Bolton:2021pey,Drewes:2022akb,Giffin:2022rei,Calderon:2022alb} can therefore be re-evaluated for studying the sensitivity of those experiments to the extra dimensions. The relation between the weak couplings of the scalar interactions with a light mediator is also possible in models of extra dimensions \cite{Maldacena:1997re, Randall:1999ee, Randall:1999vf, Appelquist:2000nn}. Next, we derive the differential cross-sections for the magnetic dipole and scalar interaction up-scattering process in the presence of extra dimensions.

\subsubsection{\label{sec:xsec} Heavy bulk neutrinos via dipole interactions in CE$\nu$NS}
\indent As pointed out earlier, the magnetic moment of massive Dirac-type
neutrinos could link the brane neutrinos $(\nu _{L})$ to all possible $n$ kk-modes of the bulk right-handed neutrinos $(\nu _{R}^{n})$ in `$d$' extra dimensions with masses
allowed by the kinematics of the process (in this case $\lesssim $ 52.9 MeV ),
through chiral interactions. 
Thus with a single photon exchange approximation with one neutrino (without
mixing), the neutrino-photon interaction in the low momentum exchange limit (%
$q^{2}\rightarrow 0$) can be written as, 
\begin{equation}
\mathcal{L}_{em}=i(\frac{2\pi \alpha \mu _{\nu }}{m_{e}})\sum_{n}{\overline{%
\nu }}_{L}\sigma _{\mu \nu }q^{\nu }{\nu _{R}^{(n)}}A^{\mu }  
\label{eq:LagMM}
\end{equation}%
\noindent where ‘$\mu _{\nu }$’ is a single neutrino magnetic dipole moment
per unit Bohr's magneton $(\mu _{B})$, ‘$\alpha$’ is the fine structure
constant, ‘$m_{e}$’ is the electron mass, ‘$A^{\mu }$’ is the electromagnetic field.
In the SM, the magnetic moment for the Dirac-type neutrino $\mu _{\nu }=\frac{3eG_{F}m_{\nu }}{8\sqrt{2}\pi ^{2}}\simeq 3.2\times
10^{-19}\mu _{B}(\frac{m_{\nu }}{1eV})$ \cite{Fujikawa:1980yx}, where ‘$e$’ is
the proton electric charge, ‘$G_{F}$’ is Fermi constant and ‘$m_{\nu }$’ is a
standard single neutrino mass. In eq. (\ref{eq:LagMM}), we sum up over the
kk-modes ‘$n$’ for an arbitrary number of large extra dimensions.

Consider a realistic case for the left-handed (or right-handed anti-neutrinos)
in the initial state, scattering off a spin-zero and spin-1/2 nuclei in the relativistic limit, which is implemented in most of the neutrino and direct dark matter detection experiments\footnote{In ref. \cite{Lindner:2016wff}, it was shown that the spin-$0$
nuclei are also good approximations of spin-$1/2$.}. Using eq. (\ref{eq:LagMM}
), we can write the transition amplitude for such a process with spin-0 target and spin-1/2 nuclei as 
\begin{equation}
i\mathcal{M}^{spin-0}=-i(\frac{2\pi \alpha \mu _{\nu }}{m_{e}q^{2}})\left[ \overline{%
\mathbf{u}}(k^{\prime })\sigma ^{\mu \nu }q_{\nu }\mathbf{u}(k)\right] \left[
Z(p_{\mu }^{\prime }+p_{\mu })F(q^{2})\right] ,  \label{eq:Amp}
\end{equation}%
and
\begin{equation}
i\mathcal{M}^{spin-1/2}=-i(\frac{2\pi \alpha \mu _{\nu }}{m_{e}q^{2}})\left[\overline{%
\mathbf{u}}(k^{\prime })\sigma ^{\mu \nu }q_{\nu }\mathbf{u}(k)\right] \left[\overline{%
\mathbf{u}}(p^{\prime })\gamma^{\mu} \mathbf{u}(p) Z F(q^{2})\right] ,  \label{eq:Amp}
\end{equation}%
where ‘$i/q^{2}$’ is the photon propagator factor, the first bracket is the
neutrino chirality-flipping tensor interaction term and the second bracket is the simplified form of the nuclear transition matrix element obtained after averaging over all the interactions between photons and quarks, ‘$Z$’ is the proton number, and ‘$F(q^{2})$’ is the nuclear form factor. Here, ‘{$\mathbf{u}(k)$’ and ‘$\mathbf{u}(k^{\prime })$’ are initial and final state neutrino spinors with four
momenta ‘$k$’ and ‘$k^{\prime }$’, respectively, ‘$p$’ and ‘$p^{\prime }$’ are four
momenta of initial and final nuclei, $\sigma ^{\mu \nu }=i/2[\gamma _{\mu }\
,\gamma _{\nu }]$ is the commutator responsible for relating left-handed
brane neutrinos with the right-handed bulk neutrinos.}

Averaging over the initial and summing over the final spin states, one can
write the averaged squared matrix element for the spin-0 and spin-1/2 cases as

\begin{align}
\langle {\vert i\mathcal{M}\vert^{2}} \rangle _{spin-0}
=\left(\frac{2\pi \alpha \mu_{\nu}Z}{m_{e}q^{2}}\right)^2
L_{0}W^{0} F^{2}(q^{2}),
\label{eq:rr0}
\end{align}

and 

\begin{align}
\langle {\vert i\mathcal{M}\vert^{2}} \rangle _{spin-1/2} =  \frac{1}{2}\left(\frac{2\pi \alpha \mu_{\nu}Z}{m_{e}q^{2}}\right)^2 L_{1/2}W^{1/2}F^{2}(q^{2}),
\label{eq:rr12}
\end{align}
where,
\begin{eqnarray}
L_{o} & = & \text{Tr}
\left[(\cancel{k^{\prime}}+M_{d})\sigma^{\mu\alpha}
q_{\alpha}P_L\cancel{k}\sigma^{\nu\beta}q_{\beta}\right], \nonumber \\
W_{o} & = & \left[%
(p_{\mu}^\prime+p_{\mu})(p_{\nu}^\prime+p_{\nu})\right], \nonumber \\
L_{1/2} & = & \text{
Tr}\left[(\cancel{k^{\prime}}+M_{d})\sigma^{\mu\alpha}q_{\alpha}P_L\cancel{k}\sigma^{\nu\beta}q_{\beta}\right],\nonumber \\
W_{1/2} & = &\text{%
Tr}\left[(\cancel{p^{\prime}}+M_{n})\gamma^{\mu}(\cancel{k}\gamma^{\nu}+M_n)\right].
\end{eqnarray}

Calculating the traces and using the relations for the four-momentum
transfer, $q=k^{\prime }-k=p-p^{\prime }$ and $q^{2}=-2M_{n}T$ with ‘$M_{n}$’
being the nuclear mass, ‘$T$’ is nuclear recoil energy, and ‘$M_d$’ is the mass of the heavy right-handed bulk neutrino, which is related to the extra dimensions by eq. (\ref{massD}). Putting the two results 
into the following general two-body scattering differential
cross-section formula (see appendix \ref{append1})
Most previous studies have considered the massless neutrino limits for the dipole interactions. An obvious question could arise whether the kinematics for the massive neutrinos in the final state could give a different cross-section. However, as we explicitly have derived (see \ref{append1}) the two-particle scattering kinematics with both incoming and outgoing massive neutrinos or with incoming massless and outgoing massive neutrinos and found that the final result remains the same as in eq. (\ref{diffcros0}), that is, independent of the masses of the
incoming and outgoing particles and only depends on the mass of the target particle, 
\begin{equation}
\left( \frac{d\sigma }{dT}\right)=\frac{\langle {\vert i\mathcal{M}\vert^{2}} \rangle}{%
32\pi M_{n}E_{\nu }^{2}}.  
\label{diffcros0}
\end{equation}%
One can get the electromagnetic differential cross section for the coherent
scattering process in the limit of relativistic incoming brane neutrinos and
massive outgoing bulk neutrinos for spin-0 and spin-1/2 target nuclei as, 
 
 \bigskip \textbf{Spin-0 case:}%
\begin{equation}
\left( \frac{d\sigma }{dT}\right) _{EM}^{spin-0}=\frac{\pi \alpha ^{2}\mu
_{\nu }^{2}}{m_{e}^{2}}\left( \frac{1}{T}-\frac{1}{E_{\nu }}+\frac{T}{%
4E_{\nu }^{2}}-\frac{M_{d}^{4}}{8M_{n}E_{\nu }^{2}T^{2}}-\frac{%
M_{d}^{2}(4E_{\nu }+2M_{n}-T)}{8M_{n}E_{\nu }^{2}T}\right) Z^{2}F^{2}(q^{2}),
\label{eq:sn0nmm}
\end{equation}

\bigskip \textbf{Spin 1/2 case:}%
\begin{equation}
\left( \frac{d\sigma }{dT}\right) _{EM}^{spin-1/2}=\frac{\pi \alpha ^{2}\mu
_{\nu }^{2}}{m_{e}^{2}}\left( \frac{1}{T}-\frac{1}{E_{\nu }}+\frac{%
M_{d}^{4}(T-M_{n})}{8M_{n}^{2}E_{\nu }^{2}T^{2}}-\frac{M_{d}^{2}(2E_{\nu
}+M_{n}-T)}{4M_{n}E_{\nu }^{2}T}\right) Z^{2}F^{2}(q^{2}).
\label{eq:sn12nmm}
\end{equation}%
The two cross-sections turn out to be different up to the following terms,

\begin{equation}
\left( \frac{d\sigma }{dT}\right) _{EM}^{spin-1/2} - \ \ \left( \frac{d\sigma }{dT}\right) _{EM}^{spin-0} = \frac{\pi  \alpha ^2 \mu_{\nu} ^2 \left(M_{d}^4+M_{d}^2 M_n T-2 M_{n}^2 T^2\right)}{8 E_{\nu}^2 m_e^2 M_{n}^2 T}.
\end{equation}
We will use (\ref{eq:sn0nmm}) in our analysis because the COHERENT experiment uses the CsI nucleus, which is spin-0. 
\subsubsection{Heavy bulk neutrinos via scalar interactions in CE$\nu$NS}
The general scalar interaction term for the neutrino vertex of the CE$\nu$NS process with the SM left-handed brane neutrinos in the initial state and massive right-handed bulk neutrinos in the final state with '$n$' kk-modes is,
\begin{equation}
\mathcal{L}_{s}=iy_{s}\sum_{n}{\overline{%
\nu }}_{L}{\nu _{R}^{(n)}}S  
\label{eq:Lags},
\end{equation}%
where ‘$S$’ is the scalar mediator, and ‘$y_s$’ is the coupling constant between the scalar field and the neutrino. Without loss of generality, we consider only scalar interactions and ignore their pseudo-scalar counterpart to focus more efficiently on the extra dimensions' contribution. Also, we assume the same coupling strength $(y_s)$ of the scalar mediators with the neutrino and target nucleus.

Following the details described in the section \ref{sec:xsec} and using the neutrino-scalar interaction term in eq. (\ref{eq:Lags}), the spin average matrix element squared of the scalar mediated CE$\nu$NS process, with a spin-0 and spin-1/2 target nuclei and approximately massless brane neutrinos and massive right-handed brane neutrinos in the final state is respectively given by
\begin{equation}
\langle {\vert i\mathcal{M}\vert^{2}} \rangle _{spin-0} =\frac{y_{s}^{4}A^{2}}{(q^{2}-M_{s}^{2})^{2}}L_{0}W^{0},
\end{equation}
and
\begin{equation}
\langle {\vert i\mathcal{M}\vert^{2}} \rangle _{spin-1/2} =\frac{y_{s}^{4}A^{2}}{2(q^{2}-M_{s}^{2})^{2}}L_{1/2}W^{1/2},
\end{equation}
where ‘$A$‘ is the total number of protons and neutrons in the target nucleus, ‘$M_s$’ is the mass of the light scalar mediator. Here the leptonic factors, ‘$L_{o}$’ and  ‘$L_{1/2}$’, and in the heavy nucleus approximation, the hadronic factors, ‘$W_{o}$’ and  ‘$W_{1/2}$’, is given by,
\begin{eqnarray}
L_{o} & = & \text{Tr}\left[ (\cancel{k^{\prime}}+m_{d})P_{L}\cancel{k}\right], \nonumber \\
W_{o} & = & \left[(p_{\mu}^\prime+p_{\mu})(p_{\nu}^\prime+p_{\nu})\right], \nonumber \\
L_{1/2} & = & \text{Tr}\left[ (\cancel{k^{\prime}}+m_{d})P_{L}\cancel{k}\right],\nonumber \\
W_{1/2} & = & \text{Tr}\left[ (\cancel{p^{\prime}}+M_{n})(\cancel{p}+M_{n})\right].
\end{eqnarray}
Using eq. (\ref{diffcros0}), the resulting cross-section in the laboratory frame for the spin-0 and spin-1/2 target nuclei are,

\bigskip \textbf{Spin-0 case:}%
\begin{equation}
\left( \frac{d\sigma }{dT}\right) _{scalar}^{spin-0}=\frac{%
y_{s}^{4}(2M_{n}+T)(2M_{n}T+M_{d}^{2})}{16\pi E_{\nu
}^{2}(2M_{n}T+M_{s}^{2})^{2}}A^{2}F^{2}(q^{2})
\label{eq:sn0scalar}
\end{equation}
and 

\bigskip \textbf{Spin-1/2 case:}%
\begin{equation}
\left( \frac{d\sigma }{dT}\right) _{scalar}^{spin-1/2}=\frac{%
y_{s}^{4}(2M_{n}+T)(2M_{n}T+M_{d}^{2})}{16\pi E_{\nu
}^{2}(2M_{n}T+M_{s}^{2})^{2}}A^{2}F^{2}(q^{2}),
\label{eq:sn12scalar}
\end{equation}
One can notice by comparing eqs. ($\ref{eq:sn0scalar}$) and ($\ref{eq:sn12scalar}$) that cross-section for the scalar interaction is independent of the spin of the target nuclei for the coherent process, unlike for the case of dipole interactions due to the neutrino magnetic moment as we saw in the previous section.

\subsection{Kinematics and size of extra dimensions}
The kinematics of different processes differ depending on the neutrinos' mass. In the 4D space-time case, when we treat neutrinos as massless, we can have the same chiral initial and final states (SM weak interaction) or opposite chiral states (dipole or scalar interactions). However, since the opposite chiral initial and final state is also possible with massive neutrinos in the final state, the mass of the final state neutrino affects the kinematics of the process. This latter possibility is relevant for the bulk neutrinos. We derive explicit expressions for all these cases in the following. Minimum neutrino energy required for the standard weak and magnetic dipole interaction process with massless neutrinos in the initial and final state to produce a nuclear recoil ‘$T$’ is given by 
\begin{equation}
E_{\nu_{min}}^{SM}=\frac{2M_n T^{2}+\sqrt{(2M_n T+T^{2})4M_n^{2}T^{2}}}{4M_n T},
\label{eqminSM}
\end{equation}%
while to produce a nuclear recoil ‘$T$’ with massive bulk neutrinos with mass, 
‘$M_{d}$’, in the final state, the above relation for a magnetic moment or scalar
interactions take the form 
\begin{equation}
E_{\nu min}^{d}=\frac{2M_n T^{2}+M_{d}^{2}T+\sqrt{%
(2M_n T+T^{2})(2M_n T+M_{d}^{2})^{2}}}{4M_n T} 
\label{eqminEM}
\end{equation}%
Likewise, the maximum recoil energy of a nucleus in the standard weak and magnetic moment with massless neutrinos is, 
\begin{equation}
T_{max}^{SM}=\frac{2M_n E_{\nu }^{2}}{2M_n E_{\nu }+M_n^{2}},  
\label{eqmaxrsm}
\end{equation}%
while the maximum recoil energy with massive right-handed bulk neutrinos in
the final state using eq. (\ref{eq:angrec}) is 
\begin{equation}
T_{max}^{d}=\frac{2M_n E_{\nu }^{2}-\left( E_{\nu }+M_n \right) M_{d}^{2}+E_{\nu }\sqrt{%
\left(2M_n(E_{\nu }+ M_{d})-M_{d}^{2}\right) \left( 2M_n(E_{\nu
}-M_{d})-M_{d}^{2}\right) }}{4M_n E_{\nu }+2M_n^{2}}.  
\label{eqmaxrem}
\end{equation}%

One can reproduce the standard electroweak case of final state massless neutrino limits, eq. (\ref{eqminSM}) from eq. (\ref{eqminEM}) and eq. (\ref{eqmaxrsm}) from eq. (\ref{eqmaxrem}), in the
limit $M_{d}\rightarrow 0$. From eq. (\ref{massD}), $M_{d}= \sqrt{m_{\nu _{R}}^{2} + \sum_{i=1...d}n_{i}^{2}/{R^{2}}}$ can be used to express the minimum neutrino energy of eq. (\ref{eqminEM}) and maximum kinetic expression of eq. (\ref{eqmaxrem}) in terms of the size of the extra dimension ‘$R$’. 

\begin{figure}\centering
\includegraphics[width=0.99\textwidth]{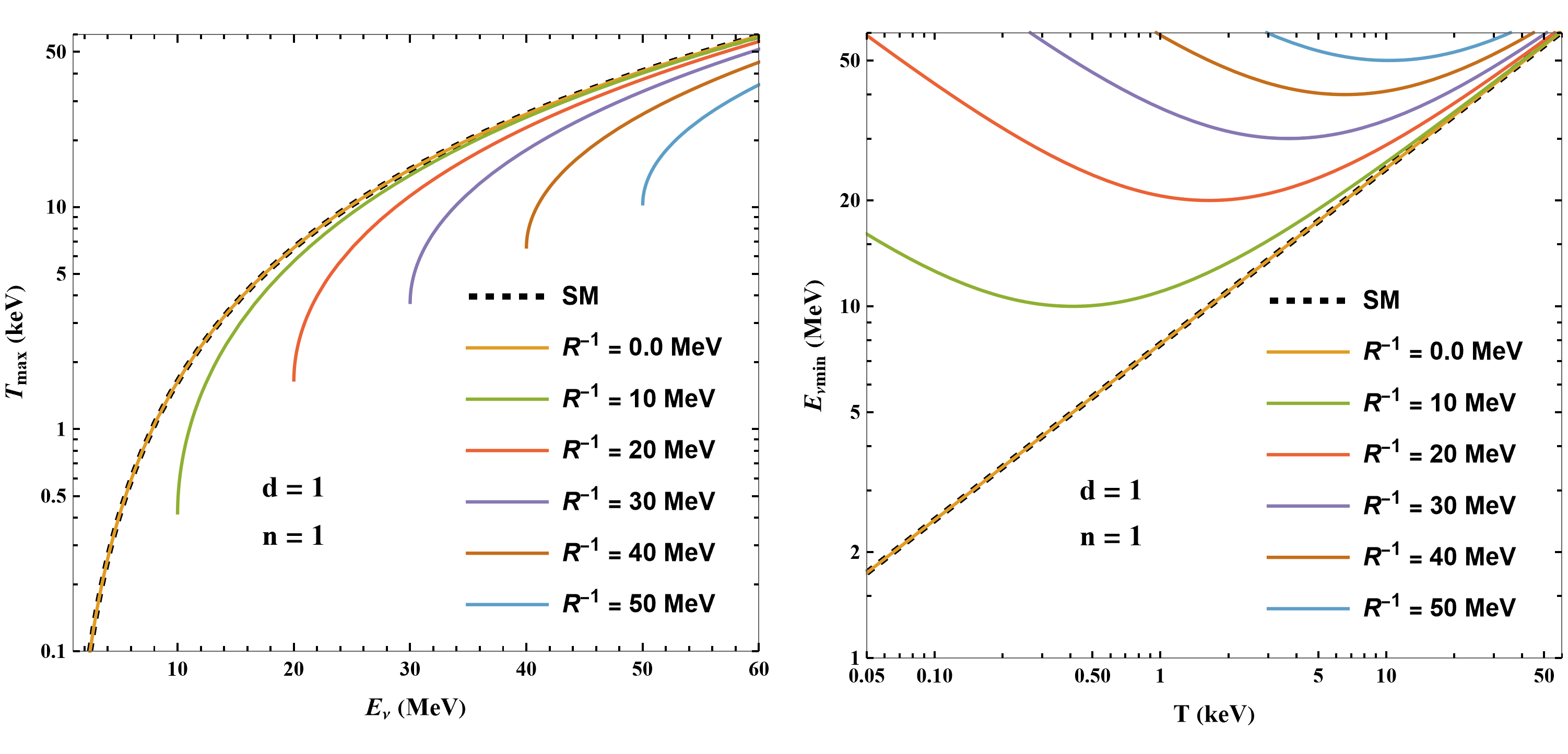}
\caption{(Left) Recoiled energy as a function of required neutrino minimum
energy for different values of the mass of bulk neutrinos allowed by the
particle kinematics for CsI target nuclei. (Right) Maximum recoiled kinetic energy for given incoming neutrino energy. The black dotted
line in both figures overlap with the $M_{d}=0$ case.}
\label{fig:kinematics}
\end{figure}

Using Eqs. (\ref{eqminEM}) and (\ref{eqmaxrem}), we show the effects of final state massive right-handed neutrinos on the correlation between the minimum neutrino energy and the recoil energy and the correlation between the maximum recoil energy and the incoming neutrino energy in fig. (\ref{fig:kinematics}). Figure (\ref{fig:kinematics}) was obtained for the ‘CsI’ nuclei, while the results scale up for other nuclei according to the nuclear mass. Notice that in fig. (\ref{fig:kinematics}), we also show the standard curves for both cases using Eqs. (\ref{eqminSM}) and  (\ref{eqmaxrsm}). They overlap with the massive neutrino cases in the limit $m_d \rightarrow 0$ in eq. (\ref{eqminEM}) and  (\ref{eqmaxrem}). We consider both cases with the simple choice of $d = 1$ and $n = 1$. The left-hand side plot shows how the extra dimension's size affects maximum nuclear recoil, while the right-hand side plot shows how the minimum neutrino required to produce a nuclear recoil depends on the size of the extra dimension. The maximum recoil energy gets smaller with smaller extra dimensions in the first case. In contrast, in the second case, the minimum energy required to produce a particular nuclear recoil increases with the smaller and smaller size of extra dimensions. Notice that the other possibilities with more than one extra dimension and higher kk-modes can also be checked along the same lines using eq. (\ref{eqminEM}) and eq. (\ref{eqmaxrem}).

\section{Event rate and Analysis}\label{sec:analysis}
We apply the framework developed above for the extra dimension to the COHERENT data of the Cesium-Iodide (CsI) target nuclear \cite{Akimov:2017ade, COHERENT:2019kwz, Akimov:2021dab, COHERENT:2021pcd}. The COHERENT detector receives a prompt signal of mono-energetic (29.8 MeV) beam of muon-neutrinos $(\nu _{\mu })$ produced in $\pi^+$ decay at rest ($\pi ^{+}\rightarrow \mu ^{+}\nu _{\mu }$) from a spallation neutron source. This is followed by a delayed emission of a broad continuous spectrum of electron-neutrinos ($\nu _{e}$) and muon-anti-neutrinos ($\bar{\nu}_{\mu })$ with energy peaks around 35 Mev and 52.9 MeV, respectively, from $\mu^+$ decays ($\mu ^{+}\rightarrow \nu_{e}e^{+}\bar{\nu}_{\mu }$) over the characteristic time scale 2.2 $\mu s$ muon lifetime. The fluxes are produced from $3.20\times 10^{23}$ protons on target from liquid mercury with an average production rate of $r=0.0848$ neutrinos for each flavor per proton \cite{Akimov:2021dab}.

For such a setup, the total number of events of the nuclear recoil in a given energy bin ‘$i$’ and neutrino flavor ‘$\alpha$’ reads
\begin{equation} 
N_{\alpha}^i=N 
\int_{T^{\prime \rm i}}^{T^{\prime \rm i+1}}\hspace{-1.5em}dT^{\prime}
\int_{0}^{T^{\rm max}}\hspace{-1.5em}dT 
\int_{E_{\nu }^{\rm min }}^{E_{\nu }^{\rm max}}\hspace{-1.3em}dE_{\nu }
\frac{d\sigma_{\alpha } }{dT}(E_{\nu },T) P_{\nu_\alpha^{(0)}\nu_\alpha^{(0)}} \frac{%
d\phi _{\nu _{\alpha }}(E_{\nu })}{dE_{\nu }}\mathcal{E} (T^\prime) G( T^\prime, T),
\label{eq:eventrt}
\end{equation}%
where $G(T^\prime, T)$ is the gamma distribution function for the detector energy resolution, $\mathcal{E} (T^\prime)$ is the detection efficiency function, $T$ and $T^\prime$ denote the nuclear recoil energy, and the reconstructed recoil energy, respectively. Both $\mathcal{E} (T^\prime)$ and  $G(T^\prime, T)$ were taken from ref. \cite{Akimov:2021dab}. $d\phi _{\nu _{\alpha }}(E_{\nu })/dE_\nu$ is the flux corresponding to the flavor ‘$\alpha$’ \cite{Khan:2021wzy}.
Here, $N=\left(2m_{\mathrm{det}}/M_{\rm CsI}\right) N_{A}$ is the total number of CsI nucleons, $m_{\mathrm{det}}=14.57$ kg, $N_{A}$ is the Avogadro's  number,  $M_{\rm CsI}$ is the molar mass of CsI, $E_{\nu }^{\min}= \sqrt{MT/2}$, $M$ is the mass of the target nucleus, $E_{\nu }^{\max }$ is the maximum neutrino energy.

In eq. (\ref{eq:eventrt}), $P_{\nu_\alpha^{(0)}\nu_\alpha^{(0)}}$ is the survival probability as given in eq. (\ref{eq:prob}) which is a function of the neutrino energy, size of extra dimension, and mass of the light neutrino in normal and inverted orderings. $d\sigma/dT(E_{\nu }, T)$ is the differential cross section of CE$\nu$NS in terms of the nuclear recoil energy, which is given by

\begin{equation}
\frac{d\sigma }{dT}\left( E_{\nu },T ; \overrightarrow{\lambda}\right) =\left( \frac{%
d\sigma }{dT}\right) _{SM}+\left( \frac{d\sigma }{dT}\right) _{LED},
\label{eq:tlxsec}
\end{equation}%
where $\overrightarrow{\lambda} \equiv (\mu_\nu, R)$ are the new physics parameters for the neutrino magnetic moment and  $\overrightarrow{\lambda} \equiv (M_s, g_s, R)$ for scalar interactions .The SM weak interaction cross-section of the CE$\nu$NS mediated by neutral current interactions reads \cite%
{Freedman:1973yd,Giunti:2007ry,Vogel:1989iv,Drukier:1983gj,Lindner:2016wff}, 
\begin{equation}
\left( \frac{d\sigma }{dT}\right) _{SM}=\frac{G_{F}^{2}M_{n}Q_{W}^{2}}{4\pi }%
\left( 1-\frac{T}{E_{\nu }}-\frac{M_{n}T}{2E_{\nu }^{2}}+\frac{T^{2}%
}{2E_{\nu }^{2}}\right) F^{2}(q^{2})\,  
\label{eq:diff}
\end{equation}%
while the LED cross-section for the neutrino magnetic moment is given in eq. (\ref{eq:sn0nmm}, \ref{eq:sn12nmm}) and for the scalar interactions is given in eq. (\ref{eq:sn0scalar}, \ref{eq:sn12scalar}). Here $Q_{W}=N-(1-4\sin ^{2}{\theta }_{W})Z$ is the weak nuclear charge, and ‘$N$’ is the neutron number. Given the form of the total differential cross-section in eq. (\ref{eq:tlxsec}) we do not expect any interference between the SM and the LED cross-sections. Notice that in eq. (\ref{eq:eventrt}), when extra dimensions through neutrino magnetic moment or scalar interactions are considered, we take $P_{\nu_\alpha^{(0)}\nu_\alpha^{(0)}}= 1$ while in case of LED through short-baseline oscillations, the extra dimensions cross-section for the neutrino magnetic moment and the scalar interaction is put equal to zero. Finally, $F(q^{2})$ in eq. \ref{eq:diff} is the nuclear form factor. We use the Klein-Nystrand
parameterization as in the following \cite{Klein:1999qj}: 
\begin{equation}
F(q^{2})=\frac{4\pi \rho _{0}}{Aq^{3}}[\sin (qR_{A})-qR_{A}\cos (qR_{A})]%
\left[ \frac{1}{1+a^{2}q^{2}}\right] .  \label{F-bessel}
\end{equation}%
Here $q^{2}=2MT$ is the momentum transfer in the scattering of neutrinos off
the nuclei, $\rho _{0}$ is the normalized nuclear density, $%
R_{A}=1.2A^{1/3}\, \mathrm{fm}$ is the nuclear radius and $a=0.7\,\mathrm{fm}
$ is the range of the Yukawa potential. 

The COHERENT measurement \cite{Akimov:2021dab} considers the energy-dependent quenching factor, $f_{\rm q}(T^{\prime})$ and measures the energy spectrum in terms of photo-electrons (p.e). For calculating the total number of events in a particular bin ‘$i$’ of photo-electrons, we use the following relation between the recoil energy and the number of photo-electrons ($N_{\rm p.e}$)
\begin{equation}\label{qf}
N_{\rm p.e.} = f_{q}(T^{\prime})\times T^{\prime}\times Y,
\end{equation}%
where $ \rm Y = 13.35 ~ \rm photons$/keV is the light yield and $f_{q}(T^{\prime})$ is taken from \cite{Akimov:2021dab}.

To fit the energy spectrum in Fig. 3 of ref. \cite{Akimov:2021dab} (also shown in fig. \ref{spectrum}) with the SM prediction and with the extra dimension models, we use the following $\chi^2$-function 
\begin{equation}
\chi ^{2}=\underset{i=2}{\overset{9}{\sum }}
\left(\frac{N_{\rm obs}^{i}-N_{\rm exp}^{i}(1+\alpha)-B^{i}(1+\beta)}
{\sigma^{i}}\right)^2
+\left( \frac{\alpha }{\sigma _{\alpha }}\right) ^{2}+\left(\frac{\beta }{
\sigma_{\beta }}\right)^{2}\,,
\label{eq:chisq}
\end{equation}%
where $N_{\rm obs}^{i}$ represents the observed events above the steady-state background in the $i$-th bin and $\sigma ^{i}$ is the relevant uncertainty \cite{Akimov:2021dab}, $N_{\rm exp}^{i}$ is the total expected events as a sum of the three neutrino flavors as given in eq.\ (\ref{eq:eventrt}) and $B^{i}$ is the sum of prior predicted beam-related neutron and the neutrino-induced neutron backgrounds in the given energy bin. The first and second penalty terms correspond to the systematic uncertainty of the signal and backgrounds where ‘$\alpha$’ and ‘$\beta$’ are the corresponding pull parameters. The uncertainty in the signal is $\sigma_{\alpha}$ = 0.127 and the uncertainty in the total background is $\sigma_{\beta}$ = 0.6 \cite{Akimov:2021dab}. The signal uncertainty includes a contribution from the quenching factor, neutrino flux, efficiency, form factor, and the light yield \cite{Akimov:2021dab}.
\section{\label{sec::Phenomenology}Phenomenology}
This section discusses the phenomenology of the three frameworks developed above and their application to the coherent elastic neutrino-nucleus scattering under certain assumptions.

\subsection{Short-baseline oscillations}
\begin{figure}[t] \centering
\includegraphics[width=1.05\textwidth,height=0.45\textwidth]{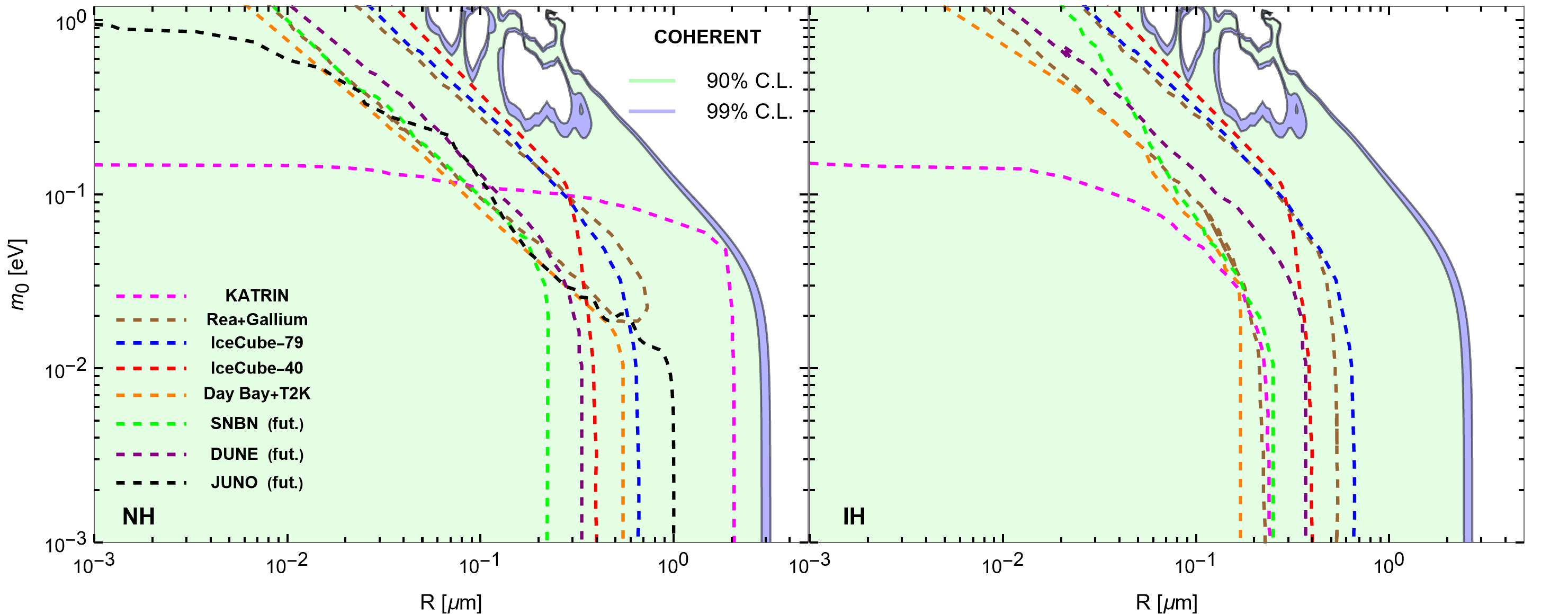}
\caption{Two dimensional 90\% and 99\% excluded regions in the ‘R-$m_o$’ space where ‘$R$’ is the LED size and $m_o = m_1$ in case of NH (Right) and  $m_o = m_1$ in case of IH (Right). For guidance, we also overlay excluded boundaries of other experiments derived in \cite{BastoGonzalez:2012me, Machado:2011jt,Machado:2011kt,DiIura:2014csa,Rodejohann:2014eka,Esmaili:2014esa,Berryman:2016szd,Adamson:2016yvy,Becker:2017ssz,Carena:2017qhd,Basto-Gonzalez:2021aus}. These results were obtained for 5 kk-modes.}
\label{eq:oscillfits}
\end{figure}

In this case, we consider the combined effect of all the three flavors of brane neutrinos ‘$\nu_{e}$’, ‘$\nu_{\mu}$’ and ‘${\overline{\nu }}_{\mu}$’ from Pion and Muon decay oscillating to the same light bulk neutrinos in the final state. We convolute the event rate in eq. (\ref{eq:eventrt}) with oscillation probability of eq. (\ref{eq:prob}) and fit the LED size ‘$R$’ and mass $m_{1}\equiv m_{0}$ for NH and $m_{3}\equiv m_{0}$ for IH using eq. (\ref{eq:chisq}). We consider five kk-modes in our analysis. The obtained exclusion regions for both mass orderings are shown in fig. \ref{eq:oscillfits}. In both cases, we also show the exclusion boundaries of the other existing and future experiments at $90\%$ C.L. \cite{BastoGonzalez:2012me, Machado:2011jt,Machado:2011kt,DiIura:2014csa,Rodejohann:2014eka,Esmaili:2014esa,Berryman:2016szd,Adamson:2016yvy,Becker:2017ssz,Carena:2017qhd,Basto-Gonzalez:2021aus}. For the COHERENT case, we show our results at $90\%$ and $99\%$ C.L. corresponding to the $\Delta\chi^2$ for two degrees of freedom. The limit on the size of large dimension is, $R \sim 3 \ \mu$m, at 90$\%$ C.L. for the absolute mass, $m_{0} \leq 3 \times 10^{-3}$ eV in case of NH. In case of IH, the limit is $R \sim 2.5 \ \mu$m, at 90$\%$ C.L. for the absolute mass, $m_{0} \leq 3 \times 10^{-3}$ eV.

Compared with other experiments, one case sees that at lower ‘$R$’ and for overall ‘$m_{0}$’, COHERENT is consistent with the other experiments. However, for the LED size $0.02 {\mu}m\leq R \leq 0.6\mu$m and $m_{0} \gtrsim 0.2$ eV the COHERENT excluded region is not smooth. One can understand this behavior by the probability function in Eqs. (\ref{eq:amplitudesur2}, \ref{eq:prob}) is highly oscillatory. In this range, the overall long-shaped oscillation regions are caused by the longer oscillation patterns above 1 MeV, as shown in fig. \ref{probb} while the additional terms in the exponents of eq. (\ref{eq:prob}) causes the inner sub-structures since they depend on ‘$m_{0}$’ and ‘$R$’. These sub-structure patterns in the form of sub-oscillations above 1 MeV for relatively different values of ‘$m_{0}$’ and ‘$R$’ can also be seen in fig. \ref{eq:oscillfits}. The overall excluded region in the case of COHERENT is relatively weaker for the current COHERENT data compared to the other experiments but competitive with KATRIN, at least in the case of NH. However, more data in the future from COHERENT and other similar experiments CE$\nu$NS experiments can certainly improve these limits \cite{Agnolet:2016zir,Akimov:2017hee,Strauss:2017cuu, Hakenmuller:2019ecb, CONUS:2020skt, CONUS:2021dwh,Bonhomme:2022lcz,CONUS:2022qbb, CONNIE:2019swq,CONNIE:2019xid,CONNIE:2021ngo,TEXONO:2020vnv,Fernandez-Moroni:2020yyl,SBC:2021yal}.

\subsection{Magnetic dipole interactions}
As discussed in section \ref{sec:xsec}, the neutrino magnetic moment changes the chirality of the incoming neutrinos in the low-energy elastic scattering processes. For left-handed neutrinos in the initial, it is possible to generate massive right-handed neutrinos (or vice versa), which could be related to the extra-dimensions through energy-momentum conservation according to eq. \ref{massD}. The heavy right-handed bulk neutrinos are free to propagate in the extra dimensions with an infinite number of kk-modes for each dimension. The maximum incoming neutrinos energy puts a natural limit on the number of extra dimensions and kk-modes.

In this case, only two parameters are involved in the analysis, the coupling of neutrino magnetic moment and the size of the extra dimension. However, the number of extra dimensions and kk-modes is arbitrary. Energy-momentum conservation of the process in 4D space-time can only constrain them. Here for convenience, we re-write the cross-section of the neutrino and spin-0 nucleus scattering used in this analysis,

\begin{equation}
\left( \frac{d\sigma }{dT}\right) _{EM}^{spin-0}=\frac{\pi \alpha ^{2}\mu
_{\nu }^{2}}{m_{e}^{2}}\left( \frac{1}{T}-\frac{1}{E_{\nu }}+\frac{T}{%
4E_{\nu }^{2}}-\frac{M_{d}^{4}}{8M_{n}E_{\nu }^{2}T^{2}}-\frac{%
M_{d}^{2}(4E_{\nu }+2M_{n}-T)}{8M_{n}E_{\nu }^{2}T}\right) Z^{2}F^{2}(q^{2}),
\label{eq:sn0nmmp}
\end{equation}
where $M_{d}$ in terms of the extra-dimensions, with the kinematical upper limit, is given by

\begin{equation}
M_{d}= \sqrt{m_{\nu _{R}}^{2} + \sum_{i=1...d}{\frac{n_{i}^{2}}{R^{2}}}} \ \ \leq E_{\nu max}.
\label{massDp}
\end{equation}

\begin{figure} \centering
\includegraphics[width=1.02\textwidth]{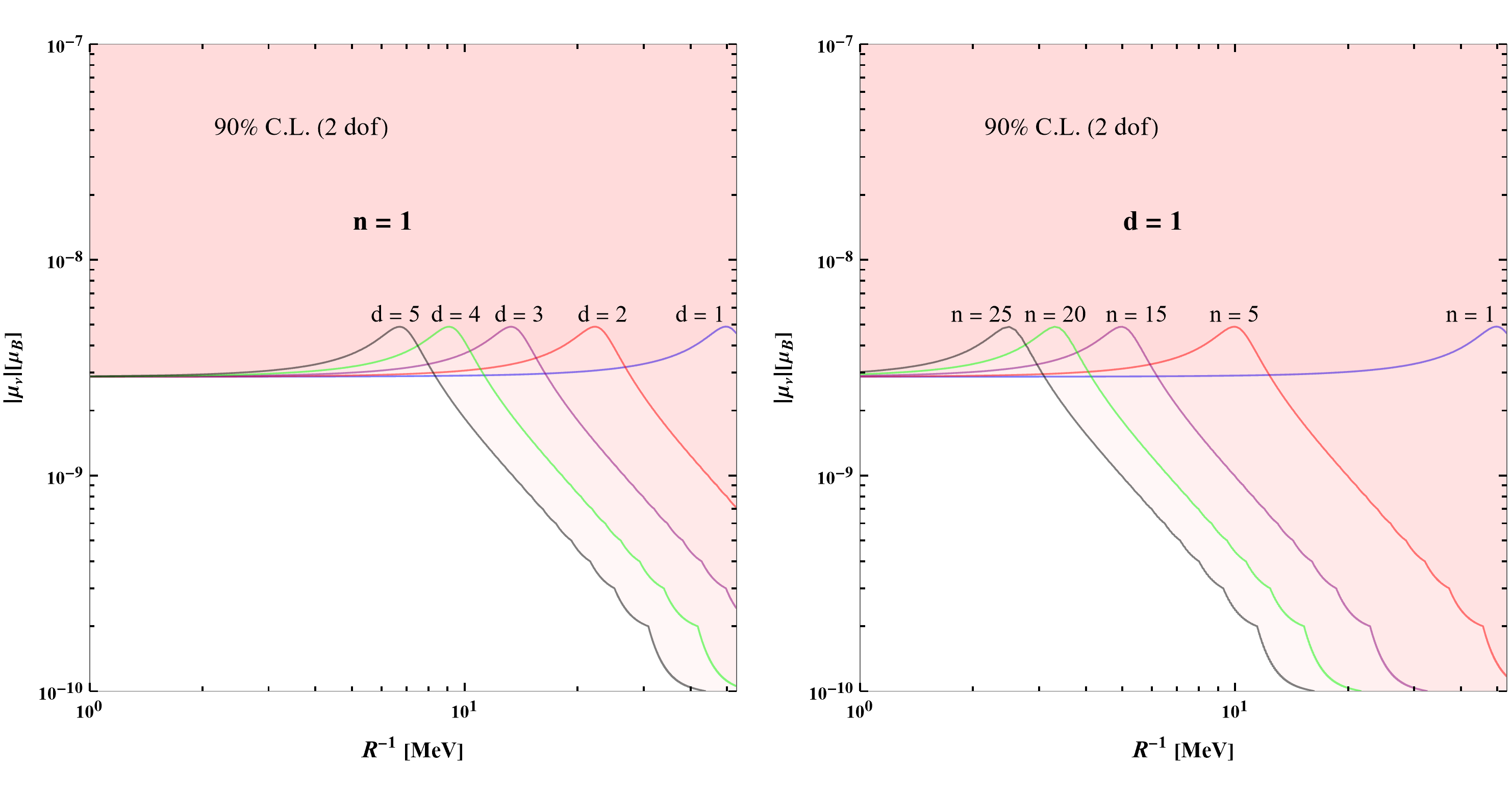}
\caption{Excluded boundary regions with 2 d.o.f at 90\% of the neutrino magnetic moment and the size of extra dimensions in the presence of heavy right-handed bulk neutrinos in the final state in the coherent scattering process. We fix the number of kk-modes to one and vary the number of extra dimensions \textbf{$(Left)$}. We fix the number of extra dimensions to one and vary the number of kk-modes \textbf{$(Right)$}.}
\label{RvsNMM}
\end{figure}

We analyze the parameter space of the neutrino effective magnetic moment and the extra dimensions in two cases. In the first case, we consider five extra dimensions and one kk-mode and the fit the dipole coupling ‘$(\mu_{\nu})$’ and the inverse size of the extra dimension ‘$(R^{-1})$’ for each of the five dimensions. In the second case, we fix the number of extra dimensions to one ‘$(d=1)$’ and consider five kk-modes, (n = 1, 5, 15, 20, 25), and fit ‘$\mu_{\nu}$’ and ‘$R^{-1}$’ in each case. We show results in fig. \ref{RvsNMM}. We show the first case on the left-handed side of fig. \ref{RvsNMM} and second case on the right-hand side of the figure. 

In both cases, we extend $(R^{-1})$ along the horizontal axis up to the kinematically allowed limit, that is, the neutrinos' maximum energy, which is about 53 MeV in the case of COHERENT. In addition to the absolute mass value of $(R^{-1})$, the number of extra dimensions and kk-modes contributes to the total mass in the final state. Therefore, in the first case (n = 1), the kinematical limit for $d = 1$ is about 53 MeV while for $d = 2, 3, 4$ and 5, it reduces for the $(R^{-1})$ along the horizontal axes. Similar, arguments apply to the second case $(d = 1)$ where the 53 MeV limit is applicable to the $n = 1$ case while for $n = 5, 15, 20$ and 25, the allowed parameter space shrinks down proportionally. However, the shrinking of the parameter spaces with increasing the number of extra dimensions (first case) or increasing the number of kk-modes (2nd case) is compensated by the smaller and smaller values of the magnetic moment couplings. The value of the dipole coupling gets smaller as the number of extra dimensions or kk-modes increases. It implies that as the number of extra dimensions or kk-modes grows, the neutrino magnetic moment and the size of the extra dimensions get smaller and smaller.

It is interesting to note that limits from the left-hand side of fig. \ref{RvsNMM} for extra dimension $d \geq 4$ and magnetic dipole coupling  $\mu_{\nu} \leq 3 \times 10^{-9} \mu_B$ are consistent with the LED model predictions, depending on the true gravitational scale. From eq. (\ref{eq:plkscale}), $R \sim 10^9$ cm for d = 4 and $R\sim 10^{11}$ cm for $d = 5$ while from the left-hand side of fig. \ref{RvsNMM},  $R\geq 2 \times 10^{-11}$ for $d = 4$ and R $\geq 3 \times 10^{-12}$ for $d = 5$. Notice that these limits correspond to $n = 1$ kk-mode. Similarly, one can obtain limits from the right-hand side of fig. \ref{RvsNMM} for $d = 1$, but with higher kk-modes. However, they are inconsistent with the LED model predictions since the $d = 1$ case is astronomically large in the LED model and inconsistent with observations. It implies that for this specific data set of the CE$\nu$NS, the fundamental gravitational energy scale to be $ \sim 1 \rm$ TeV as predicted by the LED model is allowed if one allows extra dimension $d \geq 4$ and $n = 1$.
 
\subsection{Scalar interactions}
As discussed, the scalar interactions are another possibility of the chirality flipping of the massless neutrinos in the initial state to heavy neutrinos. We have already derived the differential cross-section for $\nu-N$ coherent scattering mediated by weakly couple light scalar mediators where the outgoing neutrinos are the massive right-handed bulk neutrinos as given in eq. (\ref{eq:sn0scalar}) and eq. (\ref{eq:sn12scalar}). We assume that these neutrinos, like the case of the dipole interactions, also propagate in the extra dimensions. The differential cross-section for the spin-0 case is the following,
\begin{equation}
\left( \frac{d\sigma }{dT}\right) _{scalar}^{spin-0}=\frac{%
y_{s}^{4}(2M_{n}+T)(2M_{n}T+M_{d}^{2})}{16\pi E_{\nu
}^{2}(2M_{n}T+M_{s}^{2})^{2}}A^{2}F^{2}(q^{2})
\label{eq:sn0scalarp}
\end{equation}

where,

\begin{equation}
M_{d}= \sqrt{m_{\nu _{R}}^{2} + \sum_{i=1...d}{\frac{n_{i}^{2}}{R^{2}}}} \ \ \leq E_{\nu max}.
\label{massDp}
\end{equation}
Unlike the neutrino dipole interactions, which depend on two parameters, the scalar interactions have three free parameters, the coupling constant ‘$g_s$’, the mass of the light scalar mediator ‘$M_s$’ and the size of the extra dimension ‘$R$’. Here, we explore the conventional parameter space of ‘$g_s$’ and  ‘$M_s$’ for different choices of the size of the extra dimension, the number of extra dimensions, and the number of kk-modes.

\begin{figure} \centering
\includegraphics[width=0.8\textwidth]{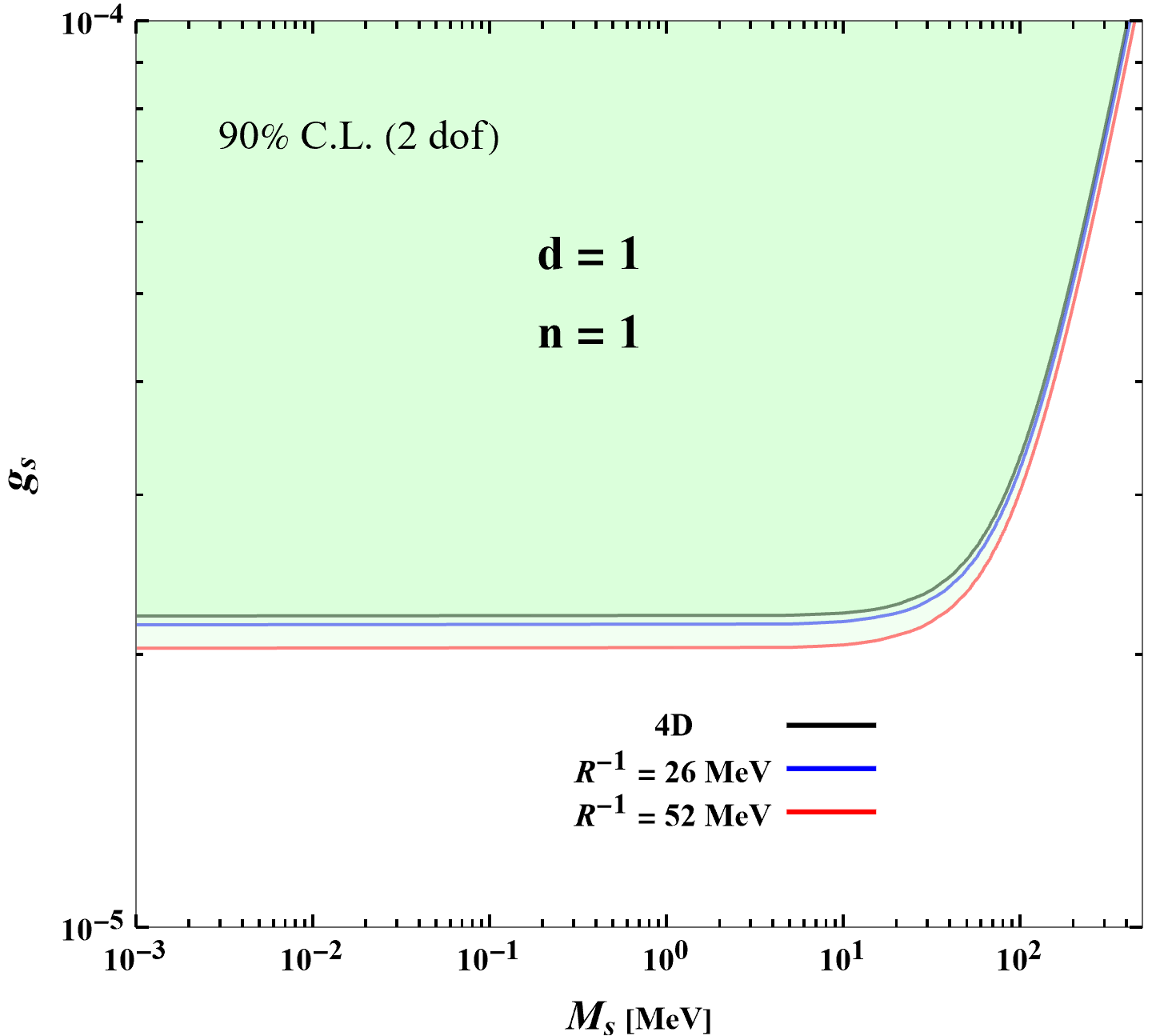}
\caption{Excluded boundary regions of the scalar coupling and mediator mass at 90\% for 2 d.o.f in the presence of heavy right-handed neutrinos connected with extra dimensions in the final state of the scattering}.
\label{rsmsvsgs}
\end{figure}

We show the results in fig. \ref{rsmsvsgs}. First, we consider $d = 1$ and $n = 1$ and fit $g_s$ and  $M_s$ for the case of massless neutrinos in the final state or with 4D (0 kk-mode) light Dirac neutrinos. As seen in fig. \ref{rsmsvsgs}, there is no distinction between the SM case where massless neutrinos are involved and the case when light Dirac type right-handed neutrinos are involved. Appreciable deviations from the 4D case occur for $R^{-1} = 26 \ \rm MeV$ and $R^{-1}= 52 \ \rm MeV$. In principle, the deviation can occur for any value of ‘$R^{-1}$’ below the maximum energy of the incoming neutrinos. For the simple case considered here, with one extra dimension and one kk-mode, the light scalar mediator couplings prefer smaller values and relatively lighter mediator masses. For more than one extra dimension and higher kk-modes, the ‘$R^{-1}$’ values will change in order to satisfy the 4D energy-momentum conservation. However, the allowed boundaries should more or less remain the same.

\begin{figure} \centering
\includegraphics[width=1\textwidth]{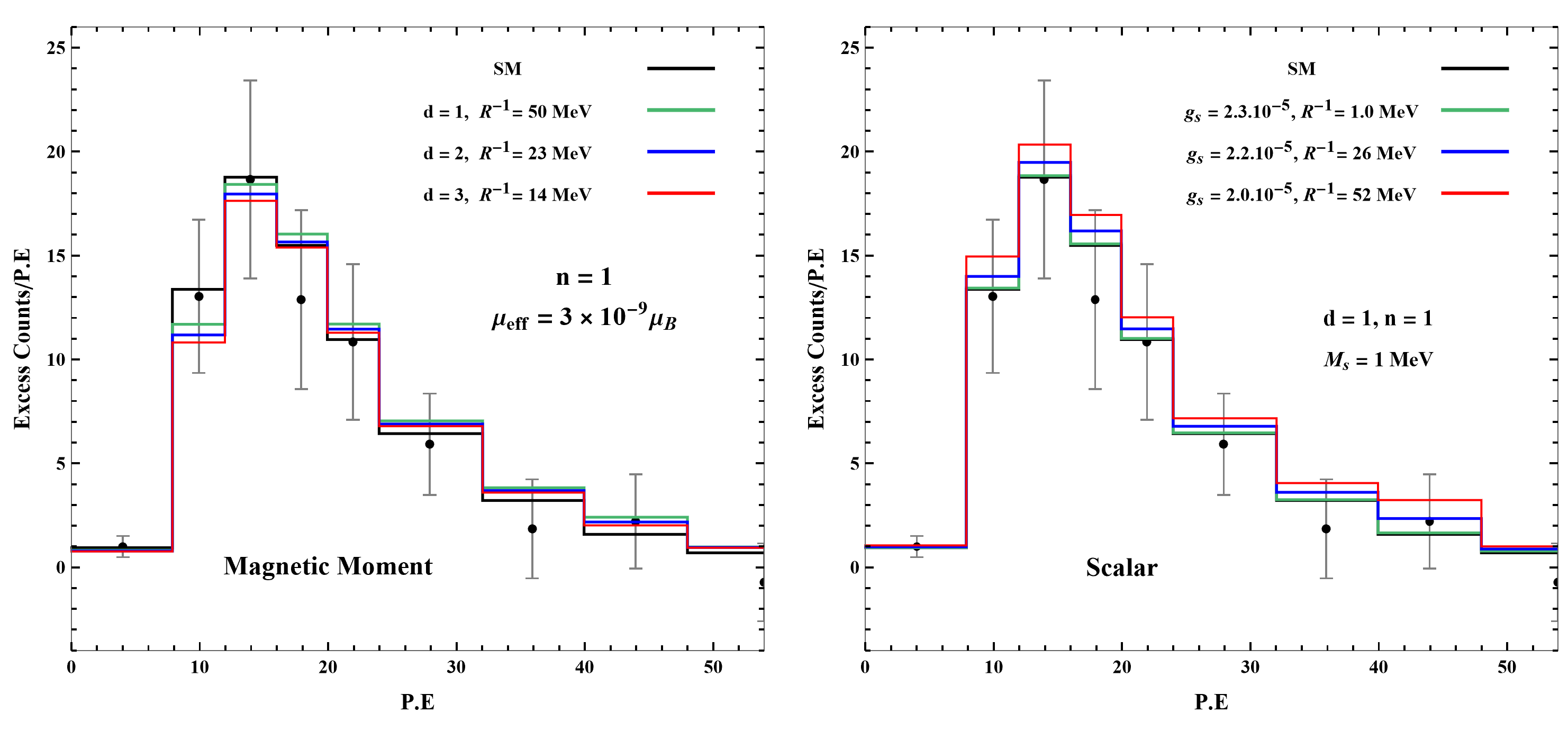}
\caption{Effects of the extra dimensions on the COHERENT recoil nuclei spectrum of the dipole interactions due to neutrino magnetic moment (left) and scalar interactions (right). We use the benchmark values as the upper limits of the magnetic moment and the size of the extra dimension (fig. \ref{RvsNMM}-left) and from the upper limits of the scalar coupling constant and the mediator mass (fig. {\ref{rsmsvsgs}}). In the case of magnetic moment, we use limits from the three extra dimensions, while in the scalar case, we use one extra dimension. In both cases, we take the one kk-mode.}
\label{spectrum}
\end{figure}

In fig. \ref{spectrum}, we show the effects of the extra dimensions on the COHERENT energy spectrum. In this figure, we have used the benchmark values obtained from the magnetic dipole interactions (fig. \ref{RvsNMM}) and scalar interactions (fig. \ref{rsmsvsgs}).
\section{\label{sec:concl} Summary and Future Outlook}
In this work, we have presented several possibilities for testing extra dimensions with
light and heavy right-handed neutrinos, called bulk neutrinos. We reviewed the mixings of light bulk and SM neutrinos in the first part. It offers an alternative phenomenology to the 4D space-time sterile neutrinos. The basis of this framework is the large extra dimension model \cite{Arkani-Hamed:2007ryu, Ibanez:2017kvh, Gonzalo:2021zsp, Harada:2022ijm, Montero:2022prj,Dvali:1999cn,Mohapatra:1999af,Mohapatra:1999zd,Ioannisian:1999cw,Mohapatra:2000wn,Barbieri:2000mg,DeGouvea:2001mz,Davoudiasl:2002fq,Bhattacharyya:2002vf}. Many authors have studied its phenomenology for various types of oscillation experiments \cite{Machado:2011jt,Machado:2011kt,BastoGonzalez:2012me,Basto-Gonzalez:2012nel,Esmaili:2014esa,Rodejohann:2014eka,DiIura:2014csa,Adamson:2016yvy,Berryman:2016szd,Carena:2017qhd,Becker:2017ssz,Basto-Gonzalez:2021aus}. We have applied this framework to the coherent elastic neutrino-nucleus scattering and used the COHERENT experimental data to constrain the size of large extra dimensions both in the case of the normal and inverted mass orderings of the SM neutrinos. We compared existing experimental limits from the other neutrino oscillation experiments and summarized the results in fig. \ref{eq:oscillfits}. The COHERENT constraint on the size of the extra dimension at 90$\%$ C.L. are, $R \sim 3 \ \mu$m (NH) and $R \sim 2.5 \ \mu$m (IH), for the neutrino absolute mass, $m_{0} \leq 3 \times 10^{-3}$ eV. The overall excluded region from COHERENT data is relatively weaker than the other limits but comparable to the bounds from the KATRIN experiment for the case of normal hierarchy.

In the second part, we have introduced a new framework for the heavy right-handed bulk neutrinos. We showed that any chirality flipping process could produce these heavy bulk neutrinos in the final state from the SM neutrinos in the initial state. The initial state neutrinos reside on the 4D brane, while the right-handed neutrinos are assumed to propagate in extra dimensions. We have shown that magnetic dipole moment or scalar interactions can produce such bulk neutrinos, where the weakly coupled scalar mediators mediate the process in the latter case. The masses of the heavy bulk neutrinos can be related to the size and number of extra dimensions through the energy-momentum relation in extra dimensions and the 4D mass. The higher dimensional momentum modes contribute to the mass of the heavy bulk neutrinos from the 4D point of view. The energy limit of the incoming neutrinos constrains the heavy bulk neutrino masses and, therefore, the size of the extra dimensions. We derived differential cross-sections for the magnetic moment and scalar interactions with spin-0 and spin-1/2 target nuclei. 

We have applied this framework to the coherent neutrino-nucleus scattering using the COHERENT data and have analyzed the two possibilities. For magnetic dipole interactions, we analyzed two cases. First, we have fixed the number of kk-modes to one and fitted the effective magnetic moment and size of the extra dimension for five extra dimensions. Second, we fixed the number of extra dimensions and changed the number of kk-modes. We have summarized the results in fig. \ref{RvsNMM}. We have shown that, depending on the true gravity scale, the constraints on the size of extra dimension ‘$R$’ are comparable to the LED prediction for $d \geq 4$ with ‘$n = 1$’ kk-mode. Following the LED model predictions, we consider the true gravitation scale of $1$ TeV. The corresponding best-fit value of the effective neutrino magnetic moment turns out to be about $3 \times 10^{-9} \mu_B$. For scalar interactions, we analyze the conventional parameter space of the coupling constant and mediator mass. We find that scalar mediator couplings prefer smaller values and relatively lighter mediator masses for the benchmark values of $d = 1, n =1$, $R^{-1} = 26$ MeV and $R^{-1} = 52$ MeV. In this case, the number of extra dimensions and kk-modes can increase, but ‘$R^{-1}$’ would decrease because of the four energy-momentum conservation. For example, the $R^{-1} = 52 $ MeV in fig \ref{rsmsvsgs} can also be obtained for $d = 1$ and $n = 5$ but with $R^{-1} = 3.5$ MeV.

The future outlook of this work is as follows. Regarding the CE$\nu$NS, we have focused on the real data collected by the COHERENT experiment. However, future experiments with accelerator and reactor neutrinos can also test the extra dimensions using the different frameworks presented here \cite{Agnolet:2016zir,Akimov:2017hee,Strauss:2017cuu, Hakenmuller:2019ecb, CONUS:2020skt, CONUS:2021dwh,Bonhomme:2022lcz,CONUS:2022qbb,CONNIE:2019swq,CONNIE:2019xid,CONNIE:2021ngo,TEXONO:2020vnv,Fernandez-Moroni:2020yyl,SBC:2021yal}. Moreover, since the size of the 4D mass of the heavy right-handed neutrinos is restricted only by the energy of the incoming neutrinos, an interesting application of this framework would be to the high energy facilities, for example, LHC \cite{Alimena:2019zri}, FASER \cite{FASER:2018bac}, MATSULA \cite{Curtin:2018mvb}, ShiP \cite{SHiP:2021nfo, Alekhin:2015byh}, Bell II \cite{Belle-II:2018jsg}, FCC \cite{FCC:2018byv}. The sensitivity studies of the heavy neutral lepton in refs.  \cite{Magill:2018jla,Shoemaker:2018vii,Bolton:2021pey,Drewes:2022akb,Giffin:2022rei,Calderon:2022alb,Abdullahi:2022jlv} can be extended by including the extra dimensions using the framework developed in this work.

We would like to point out some further physics directions of this work that might inspire future extensions. One possibility is to study the decay of the heavy extra dimension kk-modes to the SM particles and how could they change the phenomenology, for example, ref. \cite{Bolton:2021pey} has studied this possibility without extra dimensions. Another possible direction is to work on how the framework developed here applies to other models of extra dimensions such as AdS/CFT correspondence \cite{Maldacena:1997re}, warped extra dimensions \cite{Randall:1999ee, Randall:1999vf} and universal extra dimensions \cite{Appelquist:2000nn}.

\noindent
\section{\label{append}Appendix}

\subsection{\label{append1} Two particle kinematics}

In the laboratory frame, when particle ‘$B$’ is at rest in a process $%
(A+B\rightarrow A^{^{\prime }}+B),$ the standard textbook expression for
the scattering angle dependent differential cross-section, after applying
energy-momentum conservation reads as \cite{Peskin:1995ev},
\begin{equation}
\frac{d\sigma }{d\cos \theta }=-\frac{1}{32\pi m_{2}}.\frac{{|}%
\overrightarrow{{P}}{{_{3}}|}^{2}}{{|}\overrightarrow{{P}}{{_{1}}|}^{2}}.%
\frac{\overline{|i\mathcal{M}|^{2}}}{|\overrightarrow{{P}}{_{3}}%
|(E_{1}+m_{2})-E_{3}|\overrightarrow{{P}}{_{1}}|\cos \theta },
\label{kinecrss}
\end{equation}%
where $\overrightarrow{{P}}{{_{1}}}$ and $\overrightarrow{{P}}{{_{3}}}$ are
the three momenta of the initial and final state of the incident particle ($%
A\ $and $A^{^{\prime }}$), ‘$m_{2}$’ is the mass of the target particle and ‘$
E_{1}$’ and ‘$E_{3}$’ are energies of initial and final states of the incident
particle. $‘\theta’ $ is the scattering angle of the recoiled target particle with
the direction of the incident particle.

One can derive the relation between the scattering angle and recoiled electron
kinetic energy as 
\begin{equation}
d\cos \theta =-\left[ \frac{{|}\overrightarrow{{P}}{{_{3}}|}%
(E_{1}+m_{2})-E_{3}{|}\overrightarrow{{P}}{{_{1}}|}\cos \theta }{{|}%
\overrightarrow{{P}}{{_{1}}||}\overrightarrow{{P}}{{_{3}}|}^{2}}\right] dT
\label{eq:angrec}
\end{equation}%
We use this expression in eq. (\ref{kinecrss}) and obtain the well-known expression,%
\begin{equation*}
\frac{d\sigma }{dT}=\frac{1}{32\pi }.\frac{\overline{|i\mathcal{M}|^{2}}}{{|}%
\overrightarrow{{P}}{{_{1}}|}^{2}m_{2}},
\end{equation*}%
for the recoil kinetic energy-dependent differential cross-section, which we used in eq. (\ref{diffcros0}) in the limit of massless incident neutrino.

\acknowledgments
The author thanks Gia Dvali (LMU-MPI) for his valuable suggestions on several conceptual aspects of this work. The author would like to thank Werner Rodejohann (MPIK) for the initial discussions about this work and Douglas McKay (KU) for his valuable comments on the manuscript. The Alexander von Humboldt foundation financially supported this work.










\bibliographystyle{JHEP}
\bibliography{ED_DraftV2}

\end{document}